\documentclass[journal]{IEEEtran}
\IEEEoverridecommandlockouts
\usepackage{cite}
\usepackage{amsmath,amssymb,amsfonts}
\usepackage{graphicx}
\usepackage{textcomp}
\usepackage{booktabs}
\usepackage[section]{placeins}
\usepackage{xcolor}
\usepackage{xspace}
\usepackage{siunitx}
\newcommand{\ie}{\textit{i.e.}\xspace}

\begin{document}
\bstctlcite{IEEEtranControlNoDash}

\title{Fire and Forget OFDM Pilots: Transformer-Based Dynamic Wireless Channel Tracking for 5G NR}

\author{Vignesh~Natarajkumar and Lakshmi~Narasimhan~Theagarajan\thanks{The authors are with the Department of Electrical Engineering, Indian Institute of Technology Madras, Chennai 600036, India (e-mail: vigneshn@smail.iitm.ac.in; lnt@ee.iitm.ac.in).}}

\maketitle

\begin{abstract}
The availability of accurate Channel State Information (CSI) supports reliable information transfer in wireless communication systems. In time-varying fading channels, channel coefficients are estimated using pilot symbols that are transmitted every coherence interval. High-mobility wireless channels vary rapidly and require frequent pilot transmissions to maintain accurate CSI estimates. This pilot overhead significantly reduces the link throughput. To tackle this, we propose a data-driven receiver for the 5G New Radio (NR) Physical Uplink Shared CHannel (PUSCH), in which the User Equipment (UE) transmits pilots only in an initial pilot burst to the Base Station (BS). This is achieved by combining a novel channel prediction architecture with Forward Error Correction (FEC)-based refinement of channel estimates. We first consider the problem of a single-subcarrier transmission link, and then extend the design principles to a multi-subcarrier setting. We demonstrate the reduction in pilot overhead for mobilities up to \SI{400}{\km\per\hour}, and the improvement in throughput compared to both standard pilot transmission schemes and learned channel prediction baselines. 
\end{abstract}

\section{Introduction}
\label{sec:intro}
\IEEEPARstart{T}{he} fundamental problem of communication is ``that of reproducing at one point either exactly or approximately a message selected at another point''~\cite{shannon_mathematical_nodate}. To achieve this, one must understand how the wireless channel affects transmitted messages, which is characterized by the instantaneous Channel State Information (CSI). Accurate CSI estimates are essential for coherent detection of transmit data in such systems. In the 5G New Radio (NR) standard~\cite{ts38211}, where Orthogonal Frequency Division Multiplexing (OFDM) is the waveform of choice, CSI is estimated via the periodic transmission of pilot symbols, such as Demodulation Reference Signals (DMRS). 

An important feature of the time-frequency channel that OFDM deals with is its doubly selective nature: the delay spread of the multipath makes it frequency selective, and the Doppler spread caused by the relative motion of the transmitter and the receiver makes it time selective. As the selectivity of the channel increases in the time or the frequency domain, the coherence time or the coherence bandwidth decreases, respectively, requiring appropriate pilot placement and transmission. Higher relative transceiver motion increases the time selectivity of the channel, raising the pilot overhead and reducing the throughput. This is encountered in Vehicle-to-Everything (V2X) communication, drones and mobile robot networks.

With the availability of powerful hardware resources and vast amounts of wireless data, wireless communication systems, including 5G and Beyond 5G (B5G) networks, are increasingly being designed with Artificial Intelligence (AI)-native capabilities. These methods have been successfully applied to several areas previously, including physical layer design \cite{oshea_introduction_2017, dorner_deep_2018}, the creation of novel error-correcting codes \cite{hebbar_deeppolar_2024, jiang_learn_2020}, precoding \cite{lu_channel_2021}, and modeling parts of the communication process using autoencoders \cite{mattu_autoencoder_2020}. A prominent use case of AI and machine learning in the NR air interface~\cite{tr38843} is channel prediction, which is a solution to the problem of mobility management. Deep Learning (DL)-based predictors learn relevant features from collected wireless data and predict the future evolution of the channel from prior context, which may be provided by pilot transmissions. In particular, transformer-based strategies, which relate past samples to future ones at once through attention, have become popular choices for channel prediction.

Existing learned predictors, however, seldom use components of 5G NR data processing, such as error-correcting codes, to improve the quality of predictions. These components could be integral to addressing the error propagation that arises in streaming channel prediction and data transmission.

Our major contributions in this paper are:
\begin{itemize}
    \item For the problem of a single-subcarrier link, we propose a transformer-based Single-Subcarrier Predictor. We integrate this into a receiver pipeline that utilizes Decision-Directed (DD) techniques and Low-Density Parity-Check (LDPC) codes to continuously improve channel predictions, exploiting the coding gain offered by LDPC for CSI estimation.
    \item Extending this to more realistic multi-subcarrier communication, we propose a transformer-based Multi-Subcarrier Predictor for wideband OFDM channels that addresses both time and frequency selectivity. To correct our predictions and maintain a clean channel context, we also propose an auxiliary Scaler network that operates when a slot is lost.
    \item We integrate this receiver into an NR Physical Uplink Shared CHannel (PUSCH) link whose transport-block processing follows the NR specifications~\cite{ts38212, ts38214}, and demonstrate the transmission of data slots without DMRS symbols over significant durations. 
    \item We compare our model and receiver architecture with both classical pilot transmission schemes and DL-based channel prediction approaches, which shows a significant improvement in throughput over a wide range of delay and Doppler spreads. 

\end{itemize}

The rest of this paper is organized as follows. Section~\ref{sec:related} reviews related work, and Section~\ref{sec:system} describes the system model. Section~\ref{sec:single} studies the channel of a single subcarrier, and Section~\ref{sec:multi} extends it to multiple subcarriers. Section~\ref{sec:results} evaluates the OFDM receiver, and Section~\ref{sec:conclusion} concludes the paper.

\section{Related Work}
\label{sec:related}
Classical approaches to channel prediction, such as Autoregressive (AR) models~\cite{chen_channel_2021, jarinova_autoregressive_2013, duel-hallen_fading_2007}, Wiener predictors~\cite{wiener_massive_mimo}, Kalman filters, and parameter estimation methods~\cite{andersen_prediction_1999}, attempt to use the temporal correlation of the evolving channel to predict future coefficients, sometimes also considering physical wave propagation models, such as the Sum-of-Sinusoids approach. However, such methods require accurate estimates of the channel statistics. Another approach to channel prediction uses basis expansion models, which describe the time variation with a small set of basis functions. Their accuracy depends on matching the basis to mobility-induced characteristics of the channel. Such methods struggle to generalize to a large range of transceiver mobilities, resulting in suboptimal real-world performance.

This motivates the usage of learned predictors, such as random forests and k-nearest neighbors~\cite{zhang_airair_2018}.
DL-based sequence-to-sequence models~\cite{gao_fast_2023, stenhammar_comparison_2024}, such as Recurrent Neural Networks (RNNs), have been widely adopted for CSI prediction as they capture fading characteristics well. Examples of such work include Long Short-Term Memory (LSTM) networks that learn temporal correlations of fading channels~\cite{mattu_deep_2022}, hybrid Conv-CLSTM models for high-speed railways~\cite{zhou_deep-learning-based_2022}, and deep temporal Convolutional Neural Networks (CNNs) for fast-moving Clustered Delay Line (CDL) channels~\cite{gao_fast_2023}.

Transformers~\cite{vaswani_attention_2017} replace the hidden state computation of these models with attention, removing their sequential processing bottleneck. Several transformer variants have been proposed for CSI prediction~\cite{jiang_accurate_2022, zhou_transformer_2024, guo_channel_2025}, including models employing reverse~\cite{rizzello_reverse_2024} and learned~\cite{liu_channel_2025} positional encodings. 
\cite{jiang_accurate_2022} predicts the channels of several future frames in parallel with a transformer, which avoids the error propagation of sequential prediction. \cite{guo_channel_2025} proposes a CNN-Transformer architecture to predict the received Signal-to-Noise Ratio (SNR) in high-mobility multipath situations for Adaptive Modulation and Coding (AMC). \cite{zhou_transformer_2024} proposes a transformer-based predictor for CSI feedback enhancement, and evaluates it on channels measured on the Beijing-Tianjin railway line. 
\cite{nguyen_channel_2023} uses a CNN-LSTM architecture for channel estimation in a multi-user passive Reconfigurable Intelligent Surface (RIS)-aided Non-Orthogonal Multiple Access (NOMA) system. 
\cite{lin_generative_2025} compares wireless channels to natural language, and employs a Generative Pre-Trained Transformer (GPT) language model for channel prediction, and \cite{he_sensing-assisted_2025} does so in an Integrated Sensing and Communication (ISAC) setting. 

In OFDM, the channel must be predicted on multiple subcarriers, and several learned predictors are designed to tackle time and frequency selectivity jointly. For instance, STFormer~\cite{stformer} predicts the downlink channel of a Frequency Division Duplexing (FDD) link from past uplink channels, with separate attention for the temporal evolution and the spatial dependencies of the channel. \cite{zhou_timefrequency_2026} combines a frequency-domain convolution with a sparse transformer encoder for OFDM channel prediction, and validates it on CSI measured in a commercial 5G network. \cite{kronformer} is a pilot-to-prediction transformer that maps past pilots directly to the channel of future OFDM symbols in Multiple-Input Multiple-Output (MIMO) OFDM systems. \cite{lagona_p2p_2026} predicts the channel of all subcarriers in the next OFDM symbols of a single-antenna link from past pilot estimates with a transformer encoder-decoder architecture. E2ENet~\cite{e2enet} predicts the downlink channel directly from uplink pilots. In~\cite{cnnlstm_ntn}, a hybrid CNN-LSTM predicts the uplink OFDM channel of non-terrestrial links.

However, several of the channel predictors described above require the periodic transmission of reference signals to refresh their context window and limit error propagation. This means that such predictors reduce, but do not eliminate, the need for pilot symbols. A key feature of decision-directed channel estimation~\cite{porto_study_2025, mehrabi_decision_2019} is the use of correctly detected data symbols as pilots for future channel estimation, which~\cite{schafhuber_matz_2005} applied to the prediction of time-varying OFDM channels. We take this idea a step further by using data symbols that have been corrected by the error-correcting code and verified by a Cyclic Redundancy Check (CRC), to continually refresh the context window of our channel predictor, thereby eliminating the need to send pilots to restore the quality of channel tracking.
We exploit the error correction coding and interleaving present in the transceiver link to do so, and demonstrate that this indeed reduces the requirement for pilot transmissions beyond an initial pilot burst.

\section{System Model}
\label{sec:system}

\subsection{5G NR Setup}
We focus on a 5G NR PUSCH link between a single-antenna User Equipment (UE) and a single-antenna Base Station (BS). This is a wideband Single-Input Single-Output (SISO) OFDM link with $L$ useful subcarriers in which the UE transmits in consecutive uplink slots. The ideas discussed here may also be applied to other data channels, such as the downlink equivalent, the Physical Downlink Shared CHannel (PDSCH). We use the NR numerology $\mu$~\cite{ts38211}, which sets the subcarrier spacing $\Delta f = 2^{\mu} \cdot \SI{15}{\kHz}$ and the slot duration $T_{\mathrm{slot}} = 2^{-\mu}\,\si{\ms}$. A slot holds $N_{\mathrm{sym}}$ OFDM symbols, each with a normal Cyclic Prefix (CP) of duration $T_{\mathrm{CP}}$. The link occupies a channel of bandwidth $B$ and uses $N_{\mathrm{PRB}}$ Physical Resource Blocks (PRBs) of $12$ subcarriers each. A slot holds $N_{\mathrm{RE}} = 12 \cdot N_{\mathrm{PRB}} \cdot N_{\mathrm{sym}} = L N_{\mathrm{sym}}$ Resource Elements (REs).

Each slot carries a single Transport Block (TB) at a fixed Modulation and Coding Scheme (MCS). We consider MCS~$13$ ($16$-ary Quadrature Amplitude Modulation, $16$-QAM) and MCS~$7$ (Quadrature Phase Shift Keying, QPSK), with the target code rate $R$ of~\cite{ts38214}.
The TB is protected by a CRC-24A and segmented into $C$ code blocks, each with its own CRC-24B. LDPC encoding, rate matching and bit interleaving are performed as in~\cite{ts38212}. LDPC decoding is performed via the Belief Propagation (BP) algorithm~\cite{mackay_near_1996}. The bits are mapped to Gray-coded QAM symbols, where each $N_{\mathrm{cw}}$-bit LDPC codeword is mapped to $N_{\mathrm{cw}}/b$ symbols from a $2^b$-ary constellation. The resource grid in a multi-subcarrier setting is filled frequency-first. Link adaptation and Hybrid Automatic Repeat Request (HARQ) are not considered, and a TB that fails its CRC-24A is lost.

We assume perfect time and frequency synchronization and ideal hardware, \ie, no carrier frequency offset or phase noise, a single link without interference, and a receiver that knows the noise variance $N_0$. When the BS requests a new pilot burst after a lost slot, the burst starts in the next slot, \ie, the request reaches the UE without delay or error. We treat data slots without DMRS as a configuration that the BS can schedule, and do not consider a cap on the number of data REs in such slots.

\subsection{Channel Model}
We consider a time-varying frequency-selective fading channel with Root-Mean-Square (RMS) delay spread $\tau_{\mathrm{rms}}$, carrier frequency $f_c$ and maximum Doppler frequency $f_D = f_c v / c$ for a UE speed $v$. This channel is modeled using the Tapped Delay Line (TDL) model, which has a baseband representation $g(t, \tau) = \sum_{m=1}^{M} a_m[t] \cdot \delta(\tau - \tau_m)$, where $a_m[t]$ is the time-varying complex channel coefficient of the $m$-th propagation path with delay $\tau_m$. Due to relative transceiver motion, these path coefficients experience small-scale fading characterized by $f_D$. In such a setting, the frequency response of subcarrier $k$ in OFDM symbol $t$ is
\begin{equation}
H[t,k] = \sum_{m=1}^{M} a_m[t]\, e^{-j 2 \pi f_k \tau_m},
\label{eq:channel_frequency_response}
\end{equation}
where $f_k$ is the frequency of subcarrier $k$. The longest tap lies at the maximum excess delay $\tau_{\max}$, a fixed multiple of $\tau_{\mathrm{rms}}$~\cite{tr38901}. Each tap gain follows a Sum-of-Sinusoids process, and every channel realization is normalized to unit average power. 

The received symbol on RE $(t,k)$ is
\begin{equation}
y[t,k] = H[t,k]\, x[t,k] + n[t,k], \quad n[t,k] \sim \mathcal{CN}(0, N_0),
\label{eq:rx}
\end{equation}
where $x[t,k]$ is a data or pilot symbol. The channel is applied per RE and is assumed to be held constant within an OFDM symbol. We neglect the effect of Inter-Symbol Interference (ISI) as the maximum excess delay $\tau_{\max}$ of every channel we simulate is shorter than the CP duration $T_{\mathrm{CP}}$. Similarly, we neglect the effect of Inter-Carrier Interference (ICI) as the maximum Doppler frequency stays far below the subcarrier spacing, $f_D \ll \Delta f$. 

The coherence time of the channel, $T_c \approx 0.423/f_D$~\cite{rappaport_wireless_2002}, shrinks as the UE speeds up. Higher mobility therefore calls for denser pilot transmissions. Similarly, the coherence bandwidth, $B_c \approx 1/(5\tau_{\mathrm{rms}})$~\cite{rappaport_wireless_2002}, shrinks as the delay spread grows, causing the channel to change faster from one subcarrier to the next. This requires a denser allocation of pilots across the message bandwidth.

The pilot overhead is quantified by the Pilot Fraction ($\text{PF}$):
\begin{equation}
\text{Pilot Fraction, PF} = \frac{T_p}{T_p + T_d},
\label{eq:pf}
\end{equation}
where $T_p$ and $T_d$ are the total number of pilot and data symbols, respectively. As the relative mobility increases, $T_c$ shrinks, forcing a higher $\text{PF}$ and thereby reducing the throughput of the link.

\section{Single-Subcarrier Prediction}
\label{sec:single}
We first consider the channel of a single subcarrier, to identify how channel aging can be mitigated by combining a good channel predictor with standard LDPC decoding~\cite{natarajkumar_fire_2026}. Section~\ref{sec:multi} then extends these ideas to multiple subcarriers.

\subsection{Adaptations to the System Model}
\label{subsec:ss_system}
For this section, we drop the subcarrier index $k$ (\ie, we consider a flat-fading situation without frequency selectivity) and denote the received signal model as $y[t] = h[t]x[t] + n[t]$. When the channel magnitude drops significantly, the resulting deep fade drastically reduces the instantaneous received $E_b/N_0$, generating burst errors during detection and LDPC decoding. To tackle this, we assume that the transmit data is row-column interleaved, thereby introducing time diversity. At the transmitter, a block of $Q$ bits is LDPC-encoded into $C$ codewords totaling $Q/R$ bits, which are row-column interleaved and then modulated into $Q/(Rb)$ symbols, constituting one interleaved frame. The OFDM link of Section~\ref{sec:multi} does not use this interleaver. Instead, each of its LDPC codewords spans many subcarriers, so the frequency diversity of the channel and the LDPC code protect the data against a fade.

This setting differs from the OFDM link of Section~\ref{sec:multi} in $3$ further ways. First, the symbols are sent as a single-carrier stream at a symbol rate of $f_s = \SI{30}{\MHz}$, so the channel is sampled once per symbol instead of once per OFDM symbol. Second, the channel is the first tap of the TDL-A model alone, normalized to unit average power,
\begin{equation}
h[t] = a_1[t],
\end{equation}
which fades with the maximum Doppler frequency $f_D$. Third, instead of the MCS and TB processing of 5G NR, we use $16$-QAM with a fixed LDPC code rate of $R = 0.5$ and codewords of $N_{\mathrm{cw}}$ bits.

Our objective is to perform channel tracking by predicting $N$ future channel coefficients from the past $T_{\mathrm{ctx}}$ channel coefficients at each symbol step $t$. We formulate this forecasting as:
\begin{equation}
(\hat{h}[t], \dots, \hat{h}[t+N-1]) = F(h[t-T_{\mathrm{ctx}}], \dots, h[t-1]),
\label{eq:ss_multistep}
\end{equation}
where $T_{\mathrm{ctx}}$ is the context length, $N$ is the prediction horizon, and $F$ is the Single-Subcarrier Predictor we design in the following subsections.

\subsection{Data-Driven Channel Tracking Pipeline}
\label{subsec:ss_pipeline}

\begin{figure*}[t]
\centering
\includegraphics[width=\textwidth]{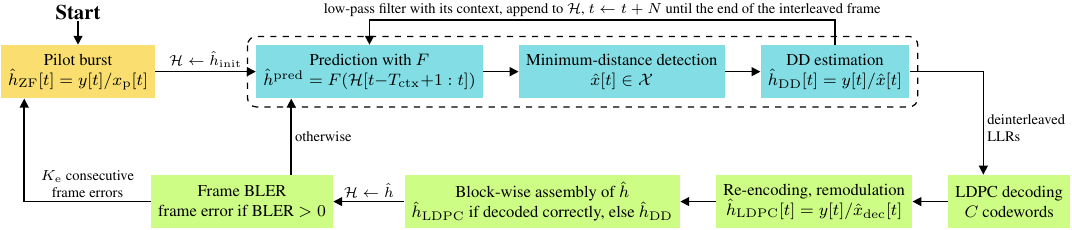}
\vspace{-2mm}
\caption{Block diagram of the proposed BLER-based channel prediction and tracking pipeline.}
\label{fig:ss_pipeline}
\vspace{-3mm}
\end{figure*}

The proposed channel tracking pipeline has a {\em Single-Subcarrier Predictor} $F$ network that predicts future channel coefficients. The Single-Subcarrier Predictor is built using the Patch Time Series Transformer (PatchTST)~\cite{nie_time_2023} as its backbone.
Let $\mathcal{H}$ denote the accumulated channel history at the receiver. The overall block diagram of the proposed pipeline is illustrated in Fig.~\ref{fig:ss_pipeline}.

\subsubsection{Initial Pilot Burst Phase}
The channel tracking begins with an initial transmission of pilots. The initial ``firing'' burst consists of $T_{\mathrm{ctx}}=N_{\text{pilot}} \allowbreak + N_{\text{data}}$ symbols with the pilot symbols uniformly spaced out among $N_{\text{data}}$ data symbols, giving a pilot fraction $\rho_p \allowbreak = N_{\text{pilot}}/(N_{\text{pilot}} + N_{\text{data}})$. The receiver performs Zero-Forcing (ZF) over the pilot indices $\Omega_{\mathrm{pilot}}$ to get the CSI estimates $\hat{h}_{\mathrm{ZF}}[t] = \frac{y[t]}{x_{\mathrm{p}}[t]}$, where $t \in \Omega_{\mathrm{pilot}}$ and $x_{\mathrm{p}}[t]$ are the known pilot symbols. These discrete estimates are linearly interpolated and low-pass filtered to obtain a continuous initial channel estimate $\hat{h}_{\mathrm{init}}[t]$ over the entire context length $T_{\mathrm{ctx}}$. This populates the initial history buffer $\mathcal{H}$ and is used to equalize the $N_{\text{data}}$ data symbols within this initial pilot burst.

\subsubsection{Pilot-Free Transmission Phase}
After the initial pilot burst, the transmitter begins the pilot-free data transmission phase. At the receiver, the most recent $T_{\mathrm{ctx}}$ samples from $\mathcal{H}$ are fed into the Single-Subcarrier Predictor $F$, which forecasts the channel states $\hat{h}^{\mathrm{pred}}[t]$ for the horizon of $N$ symbols.
Minimum-distance detection is performed on each received sample of this `chunk' as: $\hat{x}[t] = \arg\min_{x \in \mathcal{X}} \left| y[t] - \hat{h}^{\mathrm{pred}}[t]x \right|^2$, where $\mathcal{X}$ is the transmit symbol constellation. The predictions are refined in two steps:

\paragraph{Decision-Directed Refinement}
A decision-directed channel estimate $\hat{h}_{\mathrm{DD}}[t] \allowbreak = \frac{y[t]}{\hat{x}[t]}$ is obtained from each detected symbol. For correctly detected data symbols, this is akin to using them as pilot symbols. These estimates are low-pass filtered with their context and appended to $\mathcal{H}$, chunk by chunk, until the frame's $C$ LDPC blocks are processed.

\paragraph{Decoder-Aided Refinement}
Once the interleaved frame is fully received, the DD-equalized symbols are demapped to bit-level Log-Likelihood Ratios (LLRs), which are deinterleaved and segmented into LDPC codewords for BP decoding. The decoded information bits are re-encoded, re-interleaved and mapped to symbols again to generate error-corrected transmit data symbols $\hat{x}_{\mathrm{dec}}[t]$. These yield a highly accurate, LDPC-refined channel estimate $\hat{h}_{\mathrm{LDPC}}[t] = \frac{y[t]}{\hat{x}_{\mathrm{dec}}[t]}$, with which the frame is decoded a second time. As the code corrects the wrong detections of a decoded block, $\hat{h}_{\mathrm{LDPC}}[t]$ is as accurate as a pilot-based estimate there, and writing it into $\mathcal{H}$ keeps detection errors from propagating into later predictions. For each correctly decoded block, $\hat{h}_{\mathrm{LDPC}}[t]$ overwrites the DD estimates in $\mathcal{H}$.
This forms the final channel estimates $\hat{h}$ for the current frame, using which the Bit Error Rate (BER) and Block Error Rate (BLER) are computed, and the CSI tracking loop continues.

\subsubsection{CSI-Tracking Error Events}
\label{subsubsec:ss_errors}
Since $F$ is designed for all coherence times of interest, channel prediction errors occur mainly during deep fades, when the channel gain falls to nearly zero. Near deep fades, the instantaneous received $E_b/N_0$ drops, corrupting decision-directed channel estimates and symbol decisions. This also adversely affects $\mathcal{H}$ for future predictions. For high Doppler frequencies, deep fades, although relatively frequent, are short compared to the interleaved frame duration, so the time diversity introduced by interleaving and LDPC can correct such burst errors. However, for smaller Doppler frequencies, the channel is significantly more correlated, and deep fades, while rare, spread across multiple interleaved frames, leading to significant CSI tracking errors. We identify these error events by two means: (1) an oracle decision maker and (2) a BLER-based decision maker.

\paragraph{Oracle Decision Maker} We assume that an oracle gives the Normalized Mean Squared Error (NMSE) in dB for each frame's $\hat{h}$,
\begin{equation}
\mathrm{NMSE} = \frac{\sum_{t} \bigl| \hat{h}[t] - h_{\text{true}}[t] \bigr|^2}{\sum_{t} \bigl| h_{\text{true}}[t] \bigr|^2},
\label{eq:ss_nmse}
\end{equation}
where $t$ runs over the symbols of the frame and $h_{\text{true}}[t]$ is the true channel, and a CSI tracking error event is declared if the current frame's NMSE crosses a threshold.

\paragraph{BLER-Based Decision Maker} A surrogate for the CSI tracking performance is the BLER of the current frame, which can be realistically measured. An error event is declared if this BLER is nonzero in $K_{\mathrm{e}}$ consecutive frames.

\subsubsection{Pilot Restarts}
When $\mathcal{H}$ is irreparably corrupt, \ie, when a CSI tracking error event is declared, the tracker requests a new pilot burst to sanitize the channel context and restart tracking.

\subsection{Single-Subcarrier Predictor}
\label{subsec:ss_predictor}
Here, the backbone PatchTST architecture is described first, followed by the Single-Subcarrier Predictor $F$ architecture.

\subsubsection{Backbone}
The complex-valued CSI is represented as a two-channel real-valued time series, from which the backbone transformer learns the high-dimensional embeddings used for prediction.

\paragraph{Patch Embedding}
We normalize the input context with Reversible Instance Normalization (RevIN). We segment the input sequence into patches of length $P$ and stride $S$ to capture several coherence windows without drastically increasing computational complexity. The number of patches generated, where each patch ${p}_i \in \mathbb{R}^{P \times 2}$, is $N_{\mathrm{p}} = \left\lfloor \frac{T_{\mathrm{ctx}} - P}{S} \right\rfloor + 1$. The patches are encoded into a latent space via a trainable linear projection and an additive position encoding, yielding $Z_d \in \mathbb{R}^{2 \times N_{\mathrm{p}} \times d_{\mathrm{model}}}$, where $d_{\mathrm{model}}$ is the model size.

\paragraph{Transformer Encoder}
The encoder consists of a multi-head attention layer that transforms the encoder input into query, key and value matrices $Q_a = Z_d W_a^Q$, $K_a = Z_d W_a^K$, $V_a = Z_d W_a^V$, where $W_a^Q, W_a^K \in \mathbb{R}^{d_{\mathrm{model}} \times d_k}$ and $W_a^V \in \mathbb{R}^{d_{\mathrm{model}} \times d_{\mathrm{model}}}$. The scaled dot-product attention output $O_a$ is computed as $O_a = \text{Softmax}\left( \frac{Q_a K_a^T}{\sqrt{d_k}} \right) V_a$. A feed-forward network with Gaussian Error Linear Unit (GELU) activation and residual connections follows. The encoder outputs the latent representation $Z_e \in \mathbb{R}^{2 \times N_{\mathrm{p}} \times d_{\mathrm{model}}}$. We apply batch normalization and reduce overfitting via dropout in the attention and prediction layers.

\begin{figure}[t]
\centering
\includegraphics[width=\linewidth]{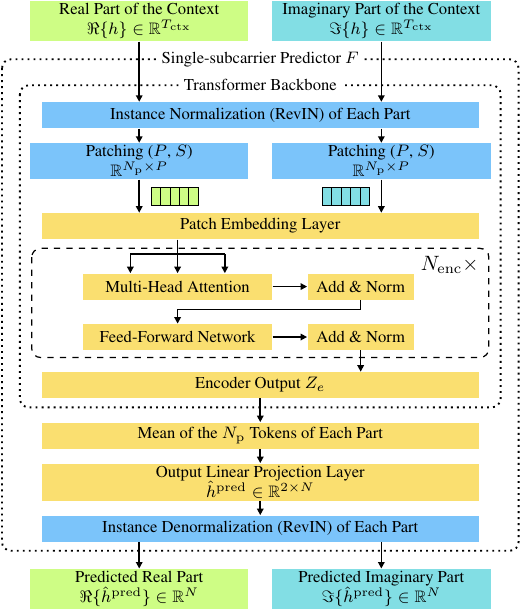}
\vspace{-3mm}
\caption{Architecture of the proposed transformer backbone and Single-Subcarrier Predictor.}
\label{fig:ss_predictor}
\vspace{-4mm}
\end{figure}

\subsubsection{Single-Subcarrier Predictor Architecture}
The input is a rolling context window of complex channel estimates $h \in \mathbb{R}^{2 \times T_{\mathrm{ctx}}}$. This context is the input to the forecaster as given on the right-hand side of~\eqref{eq:ss_multistep}. It is obtained from $\mathcal{H}$, \ie, from refined channel estimates or CSI estimates from pilot transmission.
The Single-Subcarrier Predictor network starts with the backbone, whose encoder outputs $Z_e$ are averaged over the $N_{\mathrm{p}}$ patches into one $d_{\mathrm{model}}$-dimensional vector for each of the $2$ input channels. A linear projection head and the reversal of the input normalization then map this vector to the prediction $\hat{h}^{\mathrm{pred}} \in \mathbb{R}^{2 \times N}$.
The backbone and Single-Subcarrier Predictor architectures are illustrated in Fig.~\ref{fig:ss_predictor}. Table~\ref{tab:ss_params} lists the key parameters of the link and the Single-Subcarrier Predictor, and Table~\ref{tab:ss_hparams} lists its remaining hyperparameters.

\begin{table*}[t]
\centering
\caption{System and Architecture Parameters of the Single-Subcarrier Link}
\label{tab:ss_params}
\renewcommand{\arraystretch}{1.0}
\fontsize{7.5}{9}\selectfont
\begin{tabular}{|l|c||l|c|}
\hline
\textbf{Parameter} & \textbf{Value} & \textbf{Parameter} & \textbf{Value} \\
\hline
Channel model & TDL-A & Low-pass filter cutoff frequency & \SI{80000}{\Hz} \\
Carrier frequency ($f_c$) & \SI{3.5}{\GHz} & Max. consecutive frame errors (BLER-based) ($K_{\mathrm{e}}$) & $2$ \\
LDPC code rate ($R$) & $0.5$ & Predictor context length ($T_{\mathrm{ctx}}$) & $28001$ \\
LDPC codeword length ($N_{\mathrm{cw}}$) & $6000$ bits & Predictor horizon ($N$) & $400$ \\
Blocks per interleaved frame ($C$) & $10$ & Predictor patch length ($P$) & $800$ \\
Pilot burst pilot, data symbols ($N_{\text{pilot}}$, $N_{\text{data}}$) & $4001$, $24000$ & Predictor patch stride ($S$) & $400$ \\
\hline
\end{tabular}
\vspace{-3mm}
\end{table*}

\begin{table}[!t]
\centering
\caption{Hyperparameters of the Single-Subcarrier Predictor}
\label{tab:ss_hparams}
\renewcommand{\arraystretch}{1.0}
\fontsize{7.5}{9}\selectfont
\begin{tabular}{|l|c|}
\hline
\textbf{Parameter} & \textbf{Value} \\
\hline
Predictor number of encoder layers ($N_{\mathrm{enc}}$) & $4$ \\
Predictor number of attention heads & $8$ \\
Predictor per-head dimension ($d_k$) & $128$ \\
Predictor model dimension ($d_{\mathrm{model}}$) & $1024$ \\
Predictor feed-forward dimension & $2048$ \\
Dropout probability & $0.1$ \\
Predictor batch size & $16$ \\
Maximum VAR lags & $50$ \\
\hline
\end{tabular}
\end{table}

\subsubsection{Training}
The training dataset of the Single-Subcarrier Predictor is generated with the TDL model of NVIDIA Sionna~\cite{hoydis_sionna_2023}, using the parameters in Table~\ref{tab:ss_params}. This dataset consists of $1200$ training sequences at each of $10$ different Doppler frequencies from \SI{100}{\Hz} to \SI{1400}{\Hz}, equally distributed among perfect CSI and noisy pilot-based estimates at received $E_b/N_0$ values of $-1.0$, $2.0$, and \SI{7.0}{\dB} ($16$-QAM, $R = 0.5$), a total of $12{,}000$ samples. The patch size $P$ for the Single-Subcarrier Predictor is chosen such that the coherence time $T_c$ at the maximum Doppler frequency of \SI{1400}{\Hz} spans about $11$ patches, ensuring the model observes sufficient fading in its context across a wide range of mobilities. The Single-Subcarrier Predictor is trained using Mean Squared Error (MSE): $\mathrm{MSE} = \mathbb{E} \{\| \hat{h}^{\mathrm{pred}} - h_{\text{true}}\|_2^2\}$.

The Single-Subcarrier Predictor training employs a $10$-stage curriculum~\cite{bengio_curriculum_2009}, adding one higher Doppler frequency per stage. AdamW~\cite{loshchilov_decoupled_2019} is used with a learning rate of $10^{-4}$, which is warmed up and then decayed linearly in each stage.

\subsection{Results of Single-Subcarrier Prediction}
\label{subsec:ss_results}

\begin{figure}[t]
\centering
\includegraphics[width=\linewidth]{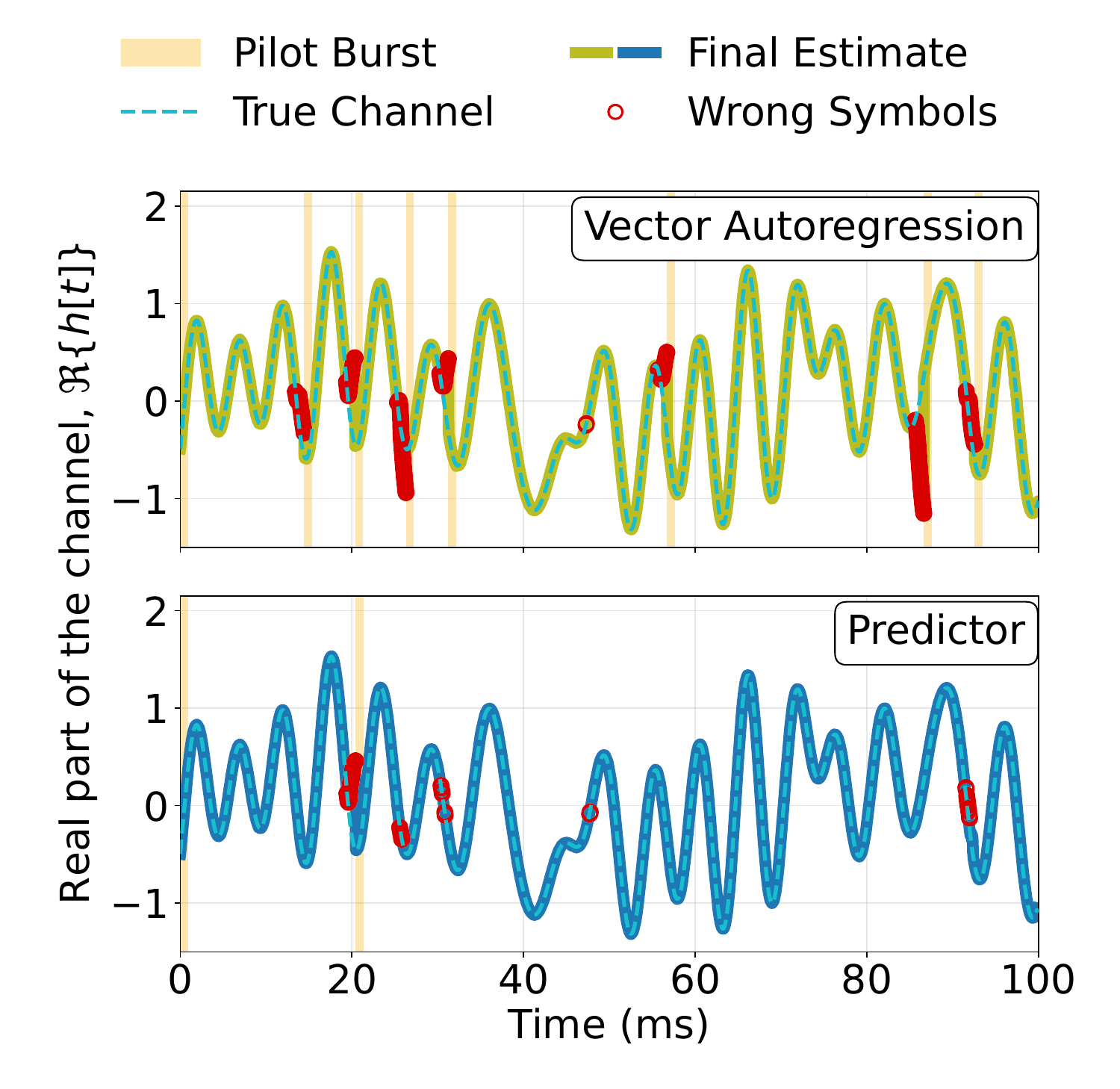}
\vspace{-4mm}
\caption{Comparison of the channel tracking capabilities of 2 channel prediction methods in the proposed receiver.}
\label{fig:ss_tracking}
\vspace{-3mm}
\end{figure}

\begin{figure}[t]
\centering
\includegraphics[width=\linewidth]{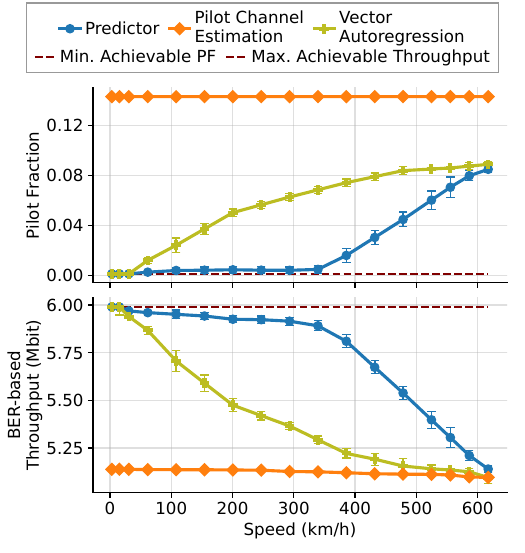}
\vspace{-5mm}
\caption{Pilot fraction (top) and BER-based throughput (bottom) of oracle-based channel tracking vs. the UE speed, $16$-QAM at a received $E_b/N_0$ of \SI{21}{\dB} and a \SI{-15}{\dB} NMSE threshold.}
\label{fig:ss_oracle}
\vspace{-3mm}
\end{figure}

\begin{figure}[t]
\centering
\includegraphics[width=\linewidth]{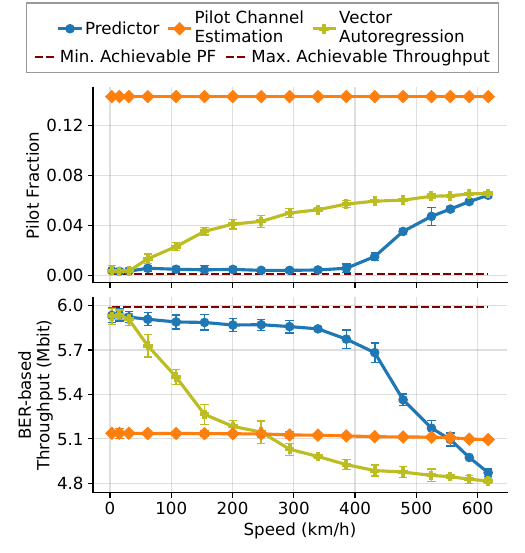}
\vspace{-5mm}
\caption{Pilot fraction (top) and BER-based throughput (bottom) of BLER-based CSI tracking vs. the UE speed for $16$-QAM at a received $E_b/N_0$ of \SI{21}{\dB}.}
\label{fig:ss_bler}
\vspace{-3mm}
\end{figure}

The proposed neural architectures were implemented using the parameters given in Tables~\ref{tab:ss_params} and~\ref{tab:ss_hparams}. All training and evaluation were performed on an NVIDIA RTX 6000 Ada Generation GPU. LDPC decoding is done with $30$ BP iterations and a flooding schedule. A maximum of $2$ LDPC decodes are required for each block. The Single-Subcarrier Predictor has $35.3$~M parameters. We evaluate the NMSE of channel prediction, the PF, and the corresponding BER over a range of UE speeds. The noise level is given as the received $E_b/N_0$ per information bit, and each channel realization is normalized to unit mean power over the transmission.

The throughput of each technique is computed over a \SI{0.1}{\s} transmission window as: $T_d\, b R (1 - \text{BER})$ bits, where $T_d$ is the number of data symbols in the window.
When only a single initial pilot burst is used for the entire \SI{0.1}{\s} transmission, the maximum achievable throughput is \SI{5.992}{\mega\bit} for $16$-QAM and the corresponding minimum achievable PF is $0.0013$. With BLER-based decisions, a burst follows only every $K_{\mathrm{e}} = 2$ frames, since the tracker is triggered when $K_{\mathrm{e}}$ consecutive frames have a nonzero BLER.

We compare (i) a conventional pilot-based CSI estimation model~\cite{dmrs}, where $1$ pilot symbol is sent with every $6$ data symbols, which corresponds to the pilot density of DMRS pos1 on a DMRS subcarrier in 5G NR~\cite{ts38211}; (ii) a CSI tracker based on Vector Autoregression (VAR)~\cite{var}; and (iii) the proposed CSI tracker with and without oracle decisions. In the figures of this section, ``Predictor'' refers to the Single-Subcarrier Predictor.

Fig.~\ref{fig:ss_tracking} provides a comparison of channel tracking capability at a relative transceiver mobility of \SI{61.71}{\km\per\hour} ($f_D = \SI{200}{\Hz}$) using BLER decisions with $16$-QAM for the VAR and the Single-Subcarrier Predictor. Clearly, the latter tracks the channel quite well. The figure also indicates that the Single-Subcarrier Predictor failed to track the channel accurately around a segment when the channel entered a deep fade. This issue, introduced in Section~\ref{subsubsec:ss_errors}, will be addressed in Section~\ref{sec:pipeline}.

{\em With oracle decisions}: Fig.~\ref{fig:ss_oracle} illustrates the PF requirements of each method, where error bars represent the standard deviation over the observations. The Single-Subcarrier Predictor drastically reduces the $0.1429$ pilot overhead of DMRS pos1 on a DMRS subcarrier.
At UE speeds up to \SI{386}{\km\per\hour} and a \SI{-15}{\dB} NMSE threshold, the Single-Subcarrier Predictor requires an average PF of just $0.0045$ for $16$-QAM.
This gives a $15.60$\% throughput improvement relative to the fixed pilot baseline. Fig.~\ref{fig:ss_oracle} also illustrates that the Single-Subcarrier Predictor, which was trained only for UE speeds up to \SI{432}{\km\per\hour}, struggles with channels that fade much faster. The VAR method is not suitable for UE speeds beyond \SI{62}{\km\per\hour}.

{\em With BLER-based decisions}: Fig.~\ref{fig:ss_bler} compares the PF requirements and the throughput when a nonzero BLER for $K_{\mathrm{e}}=2$ consecutive interleaved frames, over $C \times K_{\mathrm{e}} = 20$ blocks, is declared as an error. At UE speeds up to \SI{386}{\km\per\hour}, the Single-Subcarrier Predictor delivers $14.56$\% more throughput than the fixed pilot baseline.
Notably, these results are comparable to those of the oracle decision maker.

\section{Multi-Subcarrier Prediction}
\label{sec:multi}
We now extend the single-subcarrier receiver of Section~\ref{sec:single} to a 5G NR link, whose wideband OFDM channel varies in both time and frequency.

\subsection{Lessons from Single-Subcarrier Tracking}
\label{subsec:ofdm_system}

 In this wideband setting, we denote by $\mathbf{h}_t = [H[t, 1], \allowbreak \dots, \allowbreak H[t, L]]^T$ the channel of all subcarriers in OFDM symbol $t$, where $H[t, k]$ is defined as per~\eqref{eq:channel_frequency_response}. We formulate our objective of multi-step channel prediction, where we estimate the next $N$ channel vectors from a window of $T_{\mathrm{ctx}}$ past ones, as
\begin{equation}
\left(\hat{\mathbf{h}}_{t}, \dots, \hat{\mathbf{h}}_{t+N-1}\right) = F\left(\mathbf{h}_{t-T_{\mathrm{ctx}}}, \dots, \mathbf{h}_{t-1}\right),
\label{eq:multistep}
\end{equation}
where $F(\cdot)$ is the Multi-Subcarrier Predictor now.

Section~\ref{sec:single} showed that:
\begin{itemize}
    \item the integration of LDPC with channel prediction allows (almost) pilot-free channel tracking, especially when the parameters and training of the channel predictor are designed with the communication link in mind.
    \item the Single-Subcarrier Predictor still requests pilots when its context is irreparably corrupted, most notably when a deep fade occurs.
\end{itemize}

This section therefore aims to:
\begin{itemize}
    \item Examine how the proposed data-driven receiver works in a 5G NR-based setting, where the channel predictor now has to account for time and frequency selectivity.
    \item Address the problem of deep fades by introducing an auxiliary Scaler network that operates on lost slots and corrects prediction errors.
    \item Evaluate the resulting link performance against several factors of interest.
\end{itemize}

In this section, we choose the MCS according to the 5G NR specifications~\cite{ts38214}. We also do not perform row-column interleaving, and instead take advantage of the frequency diversity that OFDM provides to improve link robustness.

\subsection{Data-Driven Channel Tracking Pipeline}
\label{sec:pipeline}

\begin{figure*}[t]
\centering
\includegraphics[width=\textwidth]{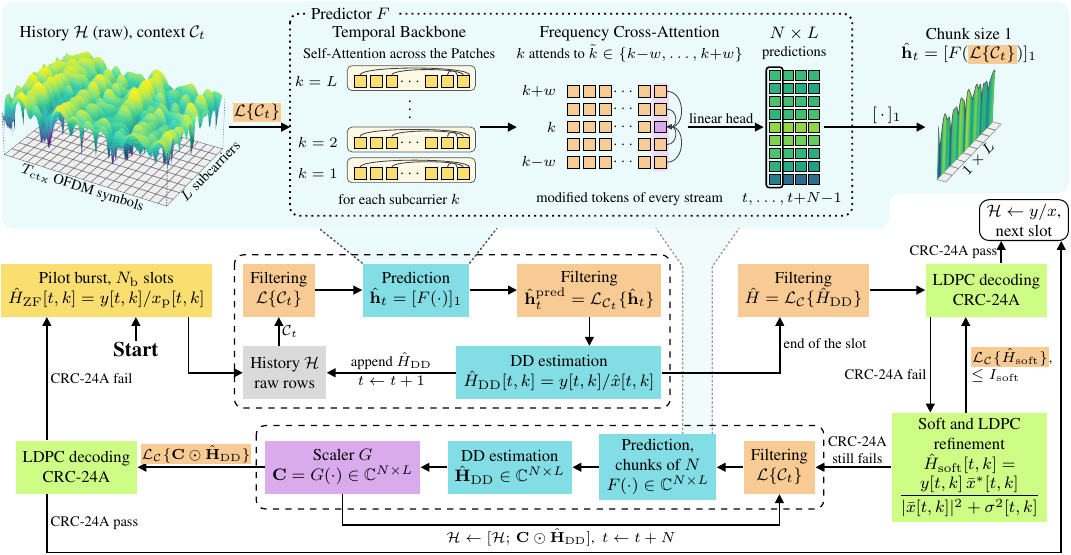}
\vspace{-2mm}
\caption{The proposed wideband channel tracking pipeline: the Multi-Subcarrier Predictor $F$ (top), the tracking of a slot one OFDM symbol at a time with decoder-aided refinement (middle), the Scaler $G$ for lost slots (bottom), and the pilot burst (left).}
\label{fig:pipeline}
\vspace{-3mm}
\end{figure*}

Fig.~\ref{fig:pipeline} shows the proposed receiver. Let $\mathcal{H}$ now be the history of all raw (unfiltered) channel estimates of the receiver, and let the context $\mathcal{C}_t$ be the $T_{\mathrm{ctx}}$ newest OFDM symbols of $\mathcal{H}$ before OFDM symbol $t$, used by the Multi-Subcarrier Predictor $F$ of~\eqref{eq:multistep} for predicting the channel. A learned Scaler $G$ repairs the predictions when a slot is lost. The context spans $T_{\mathrm{ctx}}/N_{\mathrm{sym}}$ slots. Sections~\ref{sec:predictor} and~\ref{subsec:scaler} describe the architectures of $F$ and $G$. Consider a low-pass filter $\mathcal{L}$, similar to the one used in Section~\ref{sec:single}, which is separable and applied by the Fast Fourier Transform (FFT), first along the subcarriers of each symbol (the delay domain) and then along the symbols of each subcarrier (the Doppler domain).
We denote by $\mathcal{L}_{\mathcal{C}}\{\mathbf{X}\}$ the filtering of new OFDM symbols $\mathbf{X}$ with a context $\mathcal{C}$: the receiver appends $\mathbf{X}$ to $\mathcal{C}$, filters all the OFDM symbols together with $\mathcal{L}$, and keeps only the filtered symbols of $\mathbf{X}$. The context thus smooths the new estimates, while $\mathcal{H}$ itself stays unfiltered.

\subsubsection{Initial Pilot Burst}
\label{subsec:burst}
The channel tracking begins with a pilot burst. The burst lasts $N_{\mathrm{b}}$ slots, \ie, $T_{\mathrm{ctx}} = N_{\mathrm{b}} N_{\mathrm{sym}}$ OFDM symbols, which is equal to the context window of the Multi-Subcarrier Predictor $F$. On the set $\Omega_{\mathrm{p}}$ of all pilot REs in the burst, the receiver performs ZF channel estimation $\hat{H}_{\mathrm{ZF}}[t,k] = \frac{y[t,k]}{x_{\mathrm{p}}[t,k]}, \quad (t,k) \in \Omega_{\mathrm{p}}$ where $x_{\mathrm{p}}[t,k]$ are the known pilot symbols.

The channel estimates along an OFDM symbol are interpolated to all $L$ subcarriers using the delay stage of the filter $\mathcal{L}$. Linear interpolation is then performed over time across the whole burst, extrapolating the first and last segments to the ends of the burst. This channel estimate is smoothed with $\mathcal{L}$, resulting in a pilot channel estimate.

We then equalize the data symbols in the pilot burst with this channel estimate and decode its TBs. For each TB of the burst that passes its CRC-24A check, we compute channel estimates $y[t,k]/x[t,k]$ over the TB from its re-encoded symbols $x[t,k]$ and write them into $\mathcal{H}$.

In compliance with NR processing, every slot in our pilot burst corresponds to a single-symbol DMRS with PUSCH mapping type A and configuration type~$1$~\cite[Sec.~6.4.1.1]{ts38211}. As a result, pilot symbols occupy every second subcarrier and the data fill the other half. We also allow for additional pilot symbols in every slot. As per this, the configuration pos$n$, $n \in \{0, 1, 2, 3\}$, carries $N_{\mathrm{DMRS}} = n + 1$ DMRS symbols in a TB. Each TB in our pilot burst uses DMRS pos2, where pilot symbols are found at symbol indices $2$, $7$, and $11$ within the slot. Every pilot symbol is drawn from the QPSK constellation.

\subsubsection{Pilot-Free Transmission}
\label{subsec:pilotfree}
After the burst, the transmitter sends data slots without any pilot symbols.
The Multi-Subcarrier Predictor reads the filtered context $\mathcal{L}\{\mathcal{C}_t\}$ for prediction, and the receiver keeps the first of its $N$ predicted OFDM symbols, $\hat{\mathbf{h}}_t = \left[F\left(\mathcal{L}\{\mathcal{C}_t\}\right)\right]_1$. The receiver then filters $\hat{\mathbf{h}}_t$ together with its context $\mathcal{C}_t$, which gives the predicted channel $\hat{\mathbf{h}}^{\mathrm{pred}}_t = \mathcal{L}_{\mathcal{C}_t}\{\hat{\mathbf{h}}_t\} = [\hat{H}^{\mathrm{pred}}[t,1], \dots, \hat{H}^{\mathrm{pred}}[t,L]]^T$.
The receiver performs equalization and minimum-distance detection with this predicted channel over the constellation $\mathcal{X}$. The channel estimate of the slot is then refined in two steps.

\paragraph{Decision-Directed Refinement}
For each OFDM symbol detected, we obtain a DD estimate, $\hat{H}_{\mathrm{DD}}[t,k] = {y[t,k]}/{\hat{x}[t,k]}$, which is appended to $\mathcal{H}$ before OFDM symbol $t+1$ is predicted. After the $N_{\mathrm{sym}}$ OFDM symbols of the slot, the DD estimates of the slot are filtered with the context $\mathcal{C}$ of the $T_{\mathrm{ctx}}$ OFDM symbols of $\mathcal{H}$ before the slot, $\mathcal{L}_{\mathcal{C}}\{\hat{H}_{\mathrm{DD}}\}$, and form the channel estimate $\hat{H}[t,k]$ of the slot.

\paragraph{Decoder-Aided Refinement}
The receiver equalizes the slot with $\hat{H}$ and demaps the equalized symbols to bit LLRs, which undergo LDPC decoding. Every code block is then checked by its CRC-24B and the TB by its CRC-24A. The bits and corresponding symbols of code blocks that pass their CRC-24B are treated as pilots. Once the TB passes its CRC-24A, its decoded bits are re-encoded and mapped to symbols again, which gives the transmitted symbol $x[t,k]$ on every RE of the slot, and the DD estimates of the slot in $\mathcal{H}$ are replaced with $y[t,k]/x[t,k]$. As the LDPC code has corrected the wrong detections of the slot, these estimates are as accurate as pilot-based estimates, which keeps detection errors out of the context of the next predictions.

If the TB fails, the receiver re-estimates the channel from the soft output of the decoder. From the a posteriori LLRs of the bits of each symbol, it computes the mean $\bar{x}[t,k]$ and the variance $\sigma^2[t,k]$ of the transmitted symbol $x[t,k]$. The soft-decision ZF re-estimate is
\begin{equation}
\hat{H}_{\mathrm{soft}}[t,k] = \frac{y[t,k]\, \bar{x}^*[t,k]}{|\bar{x}[t,k]|^2 + \sigma^2[t,k]} .
\label{eq:soft}
\end{equation}
The re-estimate is filtered with the same context, $\mathcal{L}_{\mathcal{C}}\{\hat{H}_{\mathrm{soft}}\}$, and the slot is equalized again. The receiver repeats this up to $I_{\mathrm{soft}}$ times, each round starting from the decoder output of the previous one. A slot counts as decoded when any attempt passes its CRC-24A, after which the DD estimates of the slot in $\mathcal{H}$ are recomputed with the known symbols as above.

\subsubsection{Prediction Scaling}
\label{subsec:rescue}
When a slot is lost despite all the channel estimation attempts of Section~\ref{subsec:pilotfree}, an underlying cause might be a channel deep fade. The corrupted DD channel estimates, after filtering, result in a complex offset from the true channel of each RE. As a result, we propose to compensate for this corruption by ``scaling'' the predicted channel estimates using a Scaler network, $G$.

The receiver tracks the lost slot's channel once again with the same context, now with the Scaler $G$ of Section~\ref{subsec:scaler} involved. For each chunk of $N$ OFDM symbols, the Multi-Subcarrier Predictor first forecasts its channel, which is refined via DD estimation. The Scaler then corrects these channel estimates by proposing a complex coefficient $C[t,k]$ for every RE of the chunk, $\mathbf{C} = G(\cdot) \in \mathbb{C}^{N \times L}$. The corrected estimates $C[t,k]\,\hat{H}_{\mathrm{DD}}[t,k]$, written $\mathbf{C} \odot \hat{\mathbf{H}}_{\mathrm{DD}}$ in Figs.~\ref{fig:pipeline} and~\ref{fig:scaler} with $\odot$ the element-wise product, are appended to $\mathcal{H}$, instead of the DD estimates $\hat{H}_{\mathrm{DD}}[t,k]$. The corrected chunks of the slot are filtered with $\mathcal{L}_{\mathcal{C}}$ and then go through the decoder-aided refinement of Section~\ref{subsec:pilotfree}.

\subsubsection{Pilot Restarts}
When every attempt at channel recovery fails, including the Scaler, a new pilot burst is requested to reset the tracking.

\subsection{Multi-Subcarrier Predictor}
\label{sec:predictor}

\begin{figure}[t]
\centering
\includegraphics[width=\linewidth]{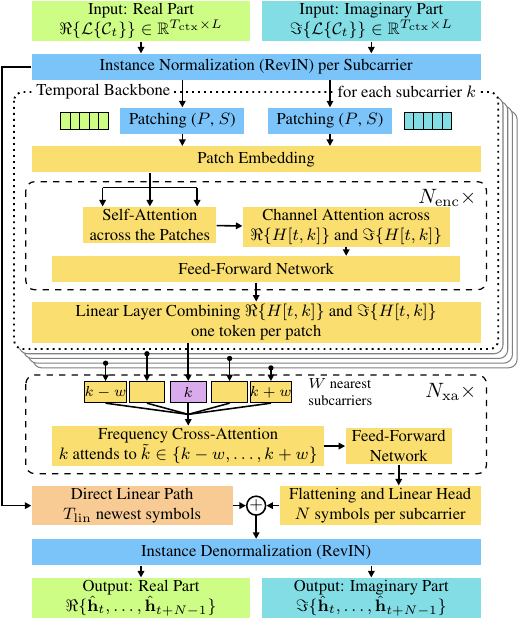}
\vspace{-2mm}
\caption{Architecture of the Multi-Subcarrier Predictor $F$.}
\label{fig:predictor}
\vspace{-3mm}
\end{figure}

\begin{table*}[!t]
\centering
\caption{System and Key Network Parameters}
\label{tab:params}
\renewcommand{\arraystretch}{1.0}
\setlength{\tabcolsep}{4.5pt}
\fontsize{7}{8.5}\selectfont
\begin{tabular}{|l|c||l|c|}
\hline
\textbf{Parameter} & \textbf{Value} & \textbf{Parameter} & \textbf{Value} \\
\hline
Channel model & TDL-A or TDL-D & Burst length ($N_{\mathrm{b}}$) & $20$ DMRS pos2 slots \\
\hline
Carrier frequency ($f_c$) & \SI{3.5}{\GHz} & Predictor context ($T_{\mathrm{ctx}}$) & $280$ symbols \\
\hline
Subcarrier spacing ($\Delta f$) & \SI{30}{\kHz} (numerology~$1$) & Predictor horizon ($N$) & $4$ symbols \\
\hline
OFDM symbols per slot ($N_{\mathrm{sym}}$) & $14$ & Patch length ($P$) & $4$ symbols \\
\hline
Bandwidth ($B$) & \SI{20}{\MHz} & Patch stride ($S$) & $2$ symbols \\
\hline
Resource blocks ($N_{\mathrm{PRB}}$) & $51$ & Cross-attention layers ($N_{\mathrm{xa}}$) & $2$ \\
\hline
RMS delay spread ($\tau_{\mathrm{rms}}$) & $10$ to \SI{200}{\ns} & Cross-attention window ($W$) & $5$ subcarriers \\
\hline
UE speed ($v$) & $10$ to \SI{400}{\km\per\hour} & Linear path input ($T_{\mathrm{lin}}$) & $128$ symbols \\
\hline
Received $E_b/N_0$ & \SI{15}{\dB} (sweep: $0$ to \SI{20}{\dB}) & Scaler history input ($T_{\mathrm{G}}$) & $28$ symbols \\
\hline
MCS~$13$: modulation, target rate ($R$) & $16$-QAM, $490/1024$ & Scaler stages ($N_{\mathrm{st}}$) & $4$ \\
\hline
MCS~$7$: modulation, target rate ($R$) & QPSK, $526/1024$ & Soft re-estimation rounds ($I_{\mathrm{soft}}$) & $3$\\
\hline
BP iterations & $30$ & Training RMS delay spreads & $10$, $50$, $100$, $150$, \SI{200}{\ns} \\
\hline
 & & Training Doppler frequencies & $100$ to \SI{1400}{\Hz} \\
\hline
\end{tabular}
\vspace{-3mm}
\end{table*}

Table~\ref{tab:params} lists the key parameters of the Multi-Subcarrier Predictor, and Table~\ref{tab:hparams} in Section~\ref{sec:results} gives further details. The Multi-Subcarrier Predictor reads the filtered context of $T_{\mathrm{ctx}}$ symbols on all $L$ subcarriers and predicts the channel for the upcoming $N$ symbols, as illustrated in Fig.~\ref{fig:predictor}. Its design follows $2$ properties of the doubly selective OFDM channel. First, the Doppler spread is similar for the narrowband channel of every subcarrier, and thus a model that captures the temporal correlation of the channel can be shared among all subcarriers. Second, due to a finite delay spread, neighboring subcarriers are correlated.

The first part of the architecture is a temporal backbone that models~\eqref{eq:ss_multistep} for each subcarrier. The complex-valued time series is normalized by RevIN and split into patches of $P$ symbols with stride $S$. It is then encoded by a PatchTST backbone, whose $N_{\mathrm{enc}}$ layers apply self-attention across the patches. Channel attention is applied across the real and imaginary parts.
A linear layer then merges the $2$ parts into one token per subcarrier and patch.

This is followed by $N_{\mathrm{xa}}$ frequency cross-attention layers which mix the tokens across neighboring subcarriers. At each patch position, the token of subcarrier $k$ attends to the tokens of the $W$ subcarriers $\tilde{k} \in \{k-w, \dots, k+w\}$, $w = (W-1)/2$, with a learned embedding of their relative offset. At the band edges, the window slides inward instead of being padded.

Finally, a linear head, shared by all subcarriers, maps the flattened tokens of each subcarrier to its channel prediction for the next $N$ OFDM symbols.
In parallel, a direct linear path maps the channel of the $T_{\mathrm{lin}}$ most recent OFDM symbols to the channel of the next $N$ OFDM symbols. Both parts of the model are trained together to produce~\eqref{eq:multistep}.

The Multi-Subcarrier Predictor is trained to minimize the MSE of the channel of the $N$ predicted symbols, over a dataset of TDL-A channels generated with Sionna over multiple Doppler frequencies and RMS delay spreads (Table~\ref{tab:params}). The training is performed in $2$ stages. First, the Multi-Subcarrier Predictor learns to forecast the clean channel from clean contexts. Then it is fine-tuned to make accurate predictions with a context of ZF channel estimates obtained from QPSK symbols, at received $E_b/N_0$ values from $0$ to \SI{30}{\dB} and also in the noise-free case. We keep the model with the best performance on validation data at a received $E_b/N_0$ of \SI{15}{\dB}.

\subsection{Multi-Subcarrier Scaler}
\label{subsec:scaler}
\begin{figure}[t]
\centering
\includegraphics[width=\linewidth]{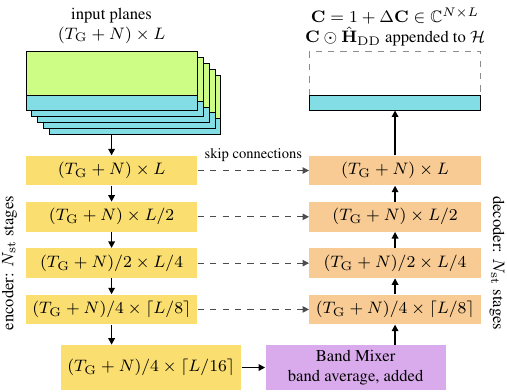}
\vspace{-2mm}
\caption{Architecture of the Scaler $G$.}
\label{fig:scaler}
\vspace{-3mm}
\end{figure}

The Scaler must propose a set of complex scaling coefficients that can adjust channel estimates to the point where a failed slot passes its CRC-24A check. For this, it must understand the degree to which the channels of different REs are distorted. As input, it is provided multiple $(T_{\mathrm{G}} + N) \times L$ grids, with real and imaginary parts of complex quantities fed as $2$ separate planes:
\begin{itemize}
    \item The channel of the most recent $T_{\mathrm{G}}$ OFDM symbols appended to the filtered prediction of the channel for the next $N$ OFDM symbols.
    \item In other input grids, the filtered prediction of the next $N$ symbols is replaced with DD channel estimates, the received power $|y|^2$ of every RE, or the unit phasor $\hat{H}_{\mathrm{DD}} \hat{H}^{\mathrm{pred}*} / |\hat{H}_{\mathrm{DD}} \hat{H}^{\mathrm{pred}}|$.
    \item A grid containing the noise level $\log_{10} (N_0/P_{\mathcal{H}})$, where $P_{\mathcal{H}}$ is the mean power of the channel over the $T_{\mathrm{G}}$ most recent OFDM symbols of $\mathcal{H}$.
    \item A grid containing the number of bits per symbol used in the transmission for that slot, replicated across the grid.
    \item Indicator grids that mark the OFDM symbols of the chunk and the position of each OFDM symbol within it.
\end{itemize}
To ensure $G$ works independently of the absolute power of the channel, the history, the prediction and the DD estimate are divided by $\sqrt{P_{\mathcal{H}}}$.

The Scaler is a convolutional encoder-decoder with skip connections, following the U-Net~\cite{ronneberger_unet_2015} architecture, as seen in Fig.~\ref{fig:scaler}. The encoder reduces the size of the grid in $N_{\mathrm{st}}$ stages and the decoder restores it in $N_{\mathrm{st}}$ stages, while the skip connections pass the features of each encoder stage to the decoder stage of the same size. At the end of the encoder, a band mixer averages the features over all subcarriers and adds a learned linear projection of this average back to every subcarrier, so that the correction of each RE also takes the state of the whole band into account. The Scaler outputs the correction $\Delta\mathbf{C}$, and $\mathbf{C} = 1 + \Delta\mathbf{C}$. As $\Delta\mathbf{C}$ is initialized to zero, an untrained Scaler leaves the DD estimates unchanged.

To train the Scaler, we generate new TDL-A channels over the same grid of Doppler frequencies and delay spreads that was used for the Multi-Subcarrier Predictor. For each channel instance, we pick the $4$ slots with the deepest fades of the received power and $1$ random slot. We do this for $2$ values of received $E_b/N_0$ taken from the sequence $6$, $9$, \dots, \SI{24}{\dB}. For each such slot in our dataset, we consider a context of channels over $T_{\mathrm{ctx}}$ OFDM symbols, composed of ZF estimates at the same noise level.

We run the receiver loop of Section~\ref{subsec:pilotfree} for these slots, tracking them in chunks of $N$ OFDM symbols as the Scaler does, and keep those whose first decoding attempt fails, resulting in $4968$ training and $1270$ validation slots. This dataset is enhanced via data augmentation. Each chunk of a kept slot is one training example. The Scaler minimizes the NMSE between the corrected estimates $C[t,k]\,\hat{H}_{\mathrm{DD}}[t,k]$ of a chunk and the true channel $H[t,k]$.

Since the corrections of the Scaler change the history that its later chunks read, we then fine-tune it in $2$ rounds of dataset aggregation~\cite{ross_reduction_2011}: each round runs the receiver with the current Scaler on the training slots and adds the chunks it produces to the training data. We keep the model with the best performance on the validation chunks.

Table~\ref{tab:params} and Table~\ref{tab:hparams} in Section~\ref{sec:results} list the parameters relevant to the Scaler.

\section{Simulations and Discussions}
\label{sec:results}

\begin{figure*}[t]
\centering
\includegraphics[width=\textwidth]{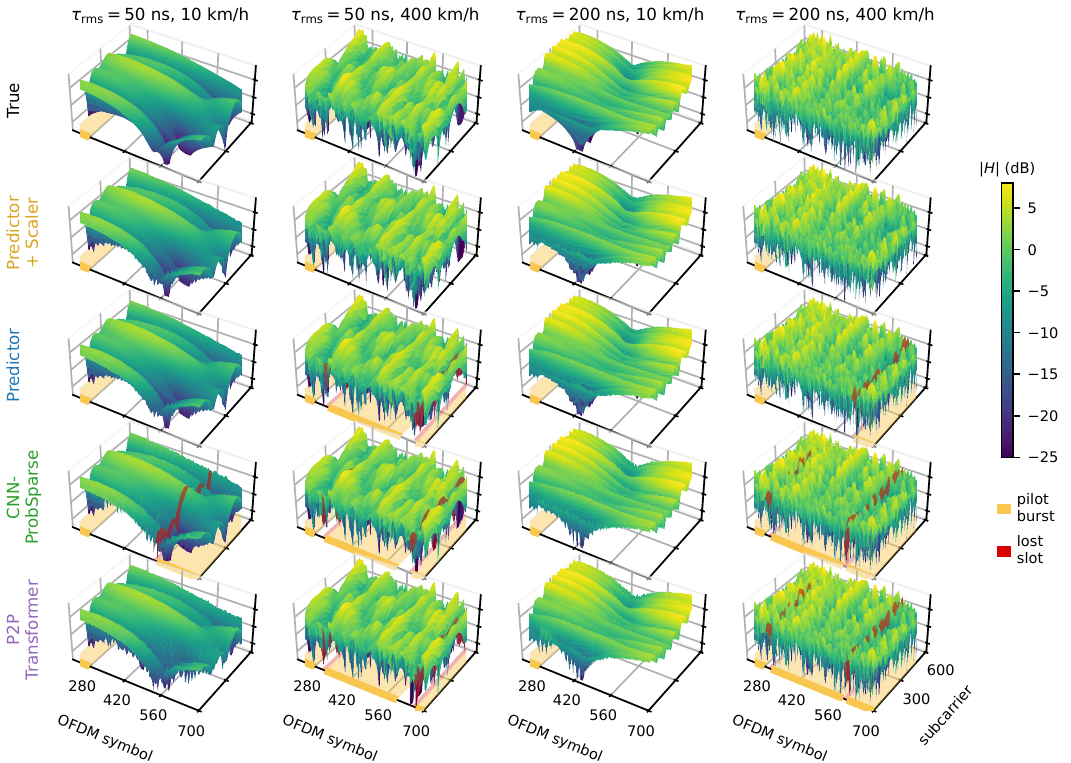}
\vspace{-6mm}
\caption{True channel magnitude (top row) and the channel with which each tracking receiver makes its decisions over $35$ slots, at a received $E_b/N_0$ of \SI{15}{\dB} using MCS~$13$, for the TDL-A channel model. Lost slots have been shaded in red, and pilot bursts in yellow.}
\label{fig:surfaces}
\vspace{-3mm}
\end{figure*}

We compare $6$ methods in total. Every method transmits exactly $1000$ slots, a duration of \SI{0.5}{\s}, including pilot bursts. Each test channel instance is normalized to unit mean power over the \SI{0.5}{\s} window. The reported values are the mean over multiple test instances, and error bars represent the standard deviation across them.
\begin{enumerate}
    \item {\em DMRS pos1 and pos2}: as discussed in Section~\ref{subsec:burst}, 5G NR-compliant transceiver links use DMRS symbols for channel estimation. We consider NR PUSCH links with every slot using DMRS pos1 or pos2.
    \item {\em CNN-ProbSparse}: our receiver without the Scaler, with the time-frequency joint predictor of~\cite{zhou_timefrequency_2026} as $F$, a convolution over subcarriers followed by a transformer encoder with ProbSparse self-attention over the newest symbols. Following the procedures of~\cite{zhou_timefrequency_2026} on our data, we widen its convolution filter to $17$ subcarriers and keep its input window of the $20$ newest OFDM symbols of the filtered context, since its Partial Autocorrelation Coefficient (PAC) criterion asks for at most $16$ symbols on our channel. We train it on the same data as our Multi-Subcarrier Predictor: first on the clean data with Adam~\cite{kingma_adam_2015} and MSE loss as per~\cite{zhou_timefrequency_2026}, then on the noisy data with our fine-tuning recipe. 
    \item {\em P2P Transformer}: our receiver without the Scaler, with the Pilot-to-Prediction (P2P) encoder-decoder transformer of~\cite{lagona_p2p_2026} as $F$. It predicts the channel of the next OFDM symbols on all subcarriers from the channel of the $30$ most recent OFDM symbols of the filtered context. It is trained and fine-tuned on our datasets with AdamW and MSE loss as per~\cite{lagona_p2p_2026}. 
    \item {\em Multi-Subcarrier Predictor}
    \item {\em Multi-Subcarrier Predictor + Scaler}
\end{enumerate}
In the figures of this section, ``Predictor'' and ``Predictor + Scaler'' refer to the Multi-Subcarrier Predictor without and with the Scaler, respectively.

\begin{table}[!t]
\centering
\caption{Hyperparameters and Training Settings of the Multi-Subcarrier Predictor, the Scaler and the Baselines}
\label{tab:hparams}
\renewcommand{\arraystretch}{1.0}
\setlength{\tabcolsep}{4.5pt}
\fontsize{7}{8.5}\selectfont
\begin{tabular}{|l|c|}
\hline
\textbf{Parameter} & \textbf{Value} \\
\hline
\multicolumn{2}{|c|}{\textit{Multi-Subcarrier Predictor}} \\
\hline
Encoder layers ($N_{\mathrm{enc}}$) & $4$ \\
Attention heads & $8$ \\
Model dimension & $256$ \\
Feed-forward dimension & $1024$ \\
Dropout & $0.1$ \\
Predictor training instances & $200$ training, $20$ validation \\
\quad per Doppler and delay spread & of $284$ symbols each \\
\hline
\multicolumn{2}{|c|}{\textit{Scaler}} \\
\hline
Scaler input planes & $13$ \\
Scaler stage widths & $32$, $64$, $96$, $128$ \\
Scaler training instances & $20$ training, $5$ validation \\
\quad per Doppler and delay spread & of $600$ symbols each \\
Scaler optimizer & AdamW, learning rate $10^{-3}$ \\
\quad schedule, batch, epochs & cosine decay, $32$, $40$ \\
\hline
\multicolumn{2}{|c|}{\textit{Baselines}} \\
\hline
Baseline~\cite{zhou_timefrequency_2026} input window & $20$ symbols \\
Baseline frequency convolution & $1 \times 17$ \\
Baseline ProbSparse layers, heads & $2$, $4$ \\
Baseline~\cite{lagona_p2p_2026} input window & $30$ symbols \\
Baseline~\cite{lagona_p2p_2026} decoder input & $10$ symbols, $4$ zero rows \\
Baseline~\cite{lagona_p2p_2026} layers, heads & $1$ encoder, $1$ decoder; $8$ \\
Baseline~\cite{lagona_p2p_2026} dimensions & $64$ (model), $256$ (feed-forward) \\
\hline
\multicolumn{2}{|c|}{\textit{Link and evaluation details}} \\
\hline
Cyclic prefix ($T_{\mathrm{CP}}$) & \SI{2.34}{\us} \\
Maximum excess delay ($\tau_{\max}$) & TDL-A: \SI{1.93}{\us} (\SI{0.48}{\us} at \SI{50}{\ns}) \\
 & TDL-D: \SI{1.25}{\us} \\
DMRS symbol positions & pos1: $\{2, 11\}$; pos2: $\{2, 7, 11\}$ \\
Code blocks per TB ($C$) at MCS~$13$ & $2$ \\
Code blocks per TB ($C$) at MCS~$7$ & $2$ (data slot); $1$ (pos1, pos2 slot) \\
LDPC base graph & $1$ \\
LLR clipping & $\pm 10$ \\
\hline
\end{tabular}
\end{table}

Table~\ref{tab:hparams} lists the remaining hyperparameters and training settings of the Multi-Subcarrier Predictor, the Scaler, and the learned baselines, together with link quantities derived from Table~\ref{tab:params}.

We report the PF and the throughput, \ie, the number of information bits in all TBs that pass their CRC-24A check within the \SI{0.5}{\s} window. The CRC bits are not counted as information bits.
We also measure the NMSE of~\eqref{eq:ss_nmse} for every slot, between a channel estimate $\hat{H}[t,k]$ and the true channel $H[t,k]$ of~\eqref{eq:channel_frequency_response} over the $N_{\mathrm{RE}} = L N_{\mathrm{sym}}$ REs of the slot:
\begin{equation}
\mathrm{NMSE} =
\frac{\sum_{t=1}^{N_{\mathrm{sym}}} \sum_{k=1}^{L} \bigl| \hat{H}[t,k] - H[t,k] \bigr|^2}
     {\sum_{t=1}^{N_{\mathrm{sym}}} \sum_{k=1}^{L} \bigl| H[t,k] \bigr|^2},
\label{eq:nmse}
\end{equation}
where $t$ runs over the OFDM symbols of the slot.

\subsection{Channel Tracking Quality}
\label{subsec:quality}
\begin{figure}[t]
\centering
\includegraphics[width=\linewidth]{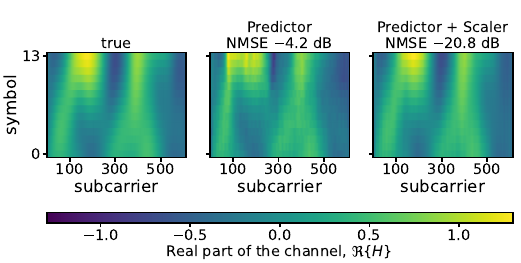}
\vspace{-5mm}
\caption{Real part of the true channel (left), the Multi-Subcarrier Predictor's estimate (middle) and the Scaler's corrected estimate (right) on a lost slot at \SI{400}{\km\per\hour} with $\tau_{\mathrm{rms}} = \SI{50}{\ns}$.}
\label{fig:scaler_effect}
\vspace{-3mm}
\end{figure}

\begin{figure}[t]
\centering
\includegraphics[width=\linewidth]{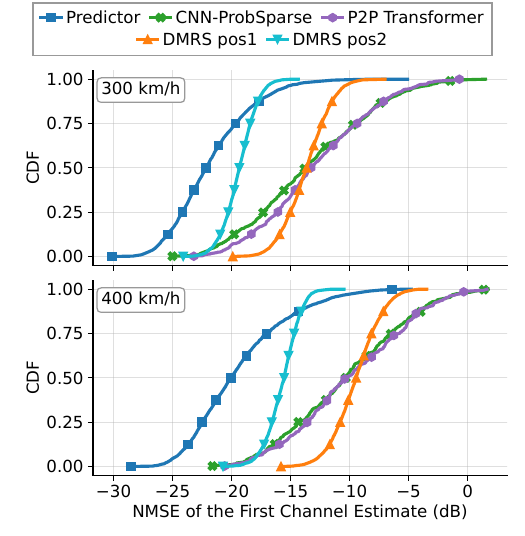}
\vspace{-5mm}
\caption{Cumulative Distribution Function (CDF) of the per-slot NMSE of the first channel estimate at UE speeds \SI{300}{\km\per\hour} (top) and \SI{400}{\km\per\hour} (bottom) at a received $E_b/N_0$ of \SI{15}{\dB} using MCS~$13$, for the TDL-A channel model with $\tau_{\mathrm{rms}} = \SI{200}{\ns}$.}
\label{fig:cdf}
\vspace{-3mm}
\end{figure}

We first evaluate how accurately the data-driven receivers track the channel.  Fig.~\ref{fig:surfaces} provides examples of channel tracking achieved by the proposed approach for a wide range of delay and Doppler spreads. With the Scaler, no slot fails during transmissions in any of the channel circumstances shown. Fig.~\ref{fig:scaler_effect} demonstrates the role of the Scaler in recovering the Multi-Subcarrier Predictor's tracking, thereby preventing the transmission of a pilot burst. The Scaler's corrected estimate has a lower NMSE than the Multi-Subcarrier Predictor's estimate.

Fig.~\ref{fig:cdf} investigates the quality of the filtered predictions of the tracking receivers at two speeds with $16$-QAM. The first estimate of a receiver is the channel with which it first detects each OFDM symbol, \ie, the filtered prediction $\hat{\mathbf{h}}^{\mathrm{pred}}_t$ for the tracking receivers and the DMRS-based channel estimate for the DMRS receivers.
DMRS pos2 has an accurate first estimate, with a median NMSE of $-19.3$ and \SI{-15.6}{\dB} at $300$ and \SI{400}{\km\per\hour}, respectively, which makes it suitable for our pilot burst at all UE speeds. Our Multi-Subcarrier Predictor is even more accurate, with medians of $-22.1$ and \SI{-20.1}{\dB}. However, $2.8$\% of the first estimates lie above \SI{-10}{\dB} at \SI{400}{\km\per\hour} ($f_D = \SI{1296}{\Hz}$), which the decoder-aided refinement of Section~\ref{sec:pipeline} has to reduce. CNN-ProbSparse and the P2P Transformer each produce a first estimate that lies above \SI{-10}{\dB} on nearly half of their data slots at \SI{400}{\km\per\hour}. This indicates that our proposed Multi-Subcarrier Predictor tracks the channel more accurately on average than both pilot channel estimation schemes and learned baselines.

\subsection{Complexity}
\label{subsec:complexity}
We first compare the sizes of the learned models. The Multi-Subcarrier Predictor has $4.64$~M trainable parameters, requiring \SI{18.6}{\mega\byte} with $32$-bit precision. An advantage of our architecture is that this requirement does not depend on, or require retraining with, an increase in the channel bandwidth. In contrast, as the dense layers of the CNN-ProbSparse and P2P Transformer are sized by the number of subcarriers, their size is dependent on the bandwidth of the link. They require $6.90$~M and $0.36$~M trainable parameters at \SI{20}{\MHz}, respectively. The Scaler has $2.62$~M trainable parameters.

\subsection{Impact of the UE Speed}
\label{subsec:mobility}

\begin{figure}[t]
\centering
\includegraphics[width=\linewidth]{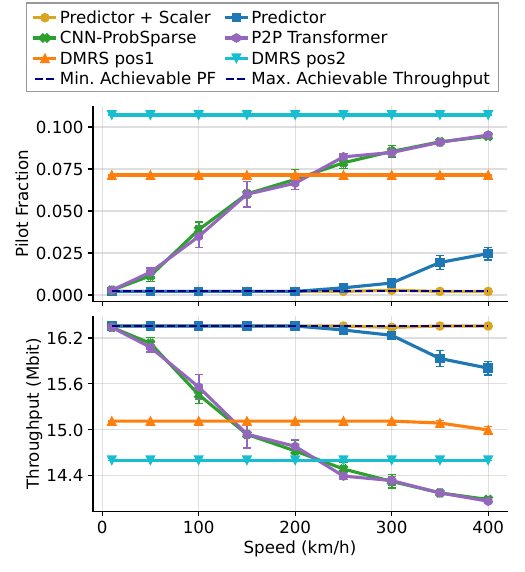}
\vspace{-5mm}
\caption{Pilot fraction (top) and correctly decoded data bits in \SI{0.5}{\s} (bottom) against the UE speed at a received $E_b/N_0$ of \SI{15}{\dB} using MCS~$13$, for the TDL-A channel model with $\tau_{\mathrm{rms}} = \SI{200}{\ns}$ ($\tau_{\max} = \SI{1.93}{\us}$).}
\label{fig:speed_mcs13}
\vspace{-3mm}
\end{figure}

\begin{figure}[t]
\centering
\includegraphics[width=\linewidth]{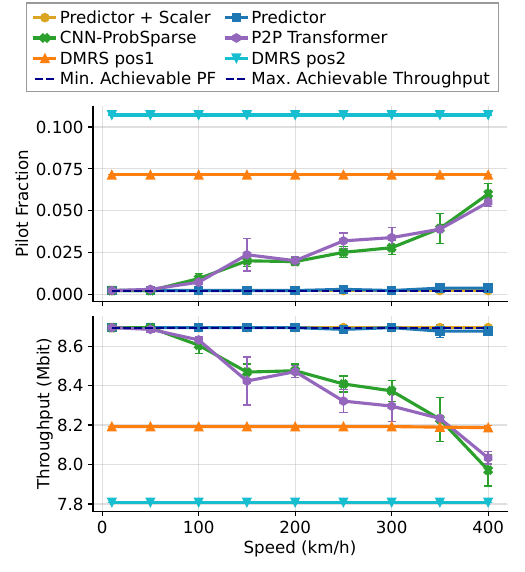}
\vspace{-5mm}
\caption{Pilot fraction (top) and correctly decoded data bits in \SI{0.5}{\s} (bottom) against the UE speed at a received $E_b/N_0$ of \SI{15}{\dB} using MCS~$7$, for the TDL-A channel model with $\tau_{\mathrm{rms}} = \SI{200}{\ns}$ ($\tau_{\max} = \SI{1.93}{\us}$).}
\label{fig:speed_mcs7}
\vspace{-3mm}
\end{figure}

We next vary the UE speed, and with it the Doppler spread of the channel.
Fig.~\ref{fig:speed_mcs13} plots the PF and throughput against the UE speed for MCS~$13$. The DMRS receivers have a fixed PF of $0.0714$ and $0.1071$, and the throughput of DMRS pos1 starts to fall once the channel ages significantly within the DMRS spacing.
DMRS pos1 loses its first TBs at \SI{350}{\km\per\hour} and $4$ to $10$ TBs per test instance at \SI{400}{\km\per\hour}, where it decodes \SI{14.996}{\mega\bit}. The learned baseline, CNN-ProbSparse, loses $16$ to $44$ TBs per test instance from \SI{100}{\km\per\hour}, and the resulting pilot overhead leads to its throughput falling below DMRS pos1 at \SI{150}{\km\per\hour} and below DMRS pos2 at \SI{250}{\km\per\hour}. The P2P Transformer decodes within \SI{0.1}{\mega\bit} of CNN-ProbSparse at every speed.

Our Multi-Subcarrier Predictor needs only the initial pilot burst up to \SI{200}{\km\per\hour} ($f_D = \SI{648}{\Hz}$), relying instead on purely data-driven tracking. At \SI{300}{\km\per\hour}, it requires $2$ to $3$ additional bursts per test instance, rising to $9$ to $12$ at \SI{400}{\km\per\hour}. The lost slots have a lower-than-average received $E_b/N_0$, indicating a potential channel fade. For example, at \SI{400}{\km\per\hour}, the received $E_b/N_0$ averaged over lost slots is \SI{12.2}{\dB}, against a \SI{15}{\dB} average. In tandem with the Scaler, however, the Multi-Subcarrier Predictor nearly always realizes pilot-free communication. Its PF stays at or close to the single-burst minimum of $0.0021$, and at \SI{400}{\km\per\hour} it decodes \SI{16.356}{\mega\bit}, the most a single burst allows. This is $9.1$\% more data bits than DMRS pos1, and Fig.~\ref{fig:speed_mcs13} shows that it also exceeds DMRS pos2 and both learned baselines.

Appropriate MCS adaptation can improve the pilot-free tracking capabilities of the proposed receiver, owing to fewer symbol detection errors and fewer erroneous DD channel estimates. This, however, usually comes at the cost of the throughput of the link. Fig.~\ref{fig:speed_mcs7} depicts this tradeoff between channel tracking and throughput. Upon switching to MCS~$7$, all three learned models demonstrate reduced PF values, with the Multi-Subcarrier Predictor performing pilot-free tracking at all UE speeds without the assistance of the Scaler. With MCS~$7$ at \SI{400}{\km\per\hour}, the Multi-Subcarrier Predictor + Scaler decodes $6.2$\% more data bits than DMRS pos1.

\subsection{Impact of the Received $E_b/N_0$}
\label{subsec:noise}

\begin{figure}[t]
\centering
\includegraphics[width=\linewidth]{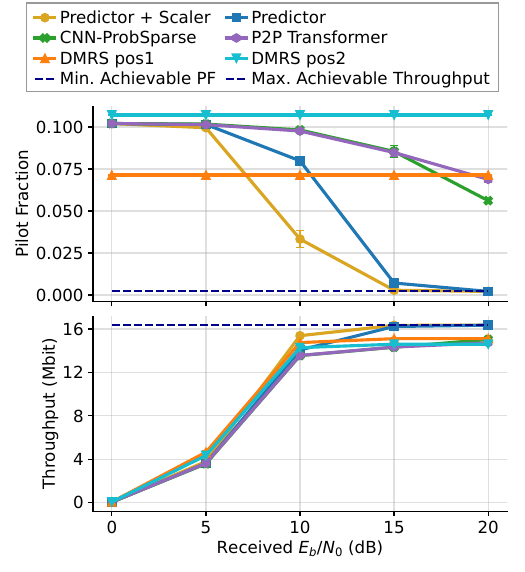}
\vspace{-5mm}
\caption{Pilot fraction (top) and correctly decoded data bits in \SI{0.5}{\s} (bottom) against the received $E_b/N_0$ at UE speed \SI{300}{\km\per\hour} using MCS~$13$, for the TDL-A channel model with $\tau_{\mathrm{rms}} = \SI{200}{\ns}$ ($\tau_{\max} = \SI{1.93}{\us}$).}
\label{fig:snr}
\vspace{-3mm}
\end{figure}

\begin{figure}[t]
\centering
\includegraphics[width=\linewidth]{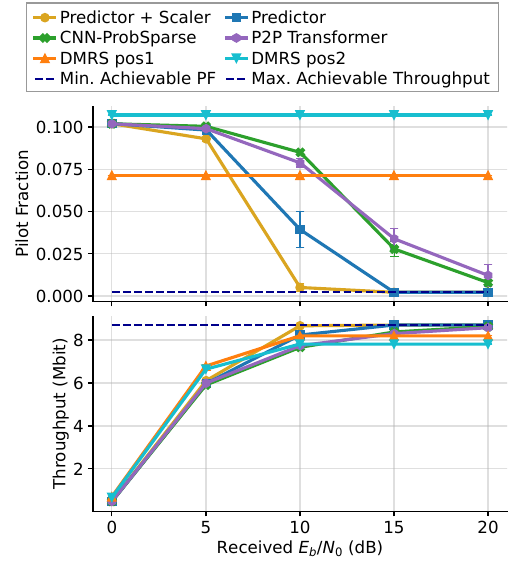}
\vspace{-5mm}
\caption{Pilot fraction (top) and correctly decoded data bits in \SI{0.5}{\s} (bottom) against the received $E_b/N_0$ at UE speed \SI{300}{\km\per\hour} using MCS~$7$, for the TDL-A channel model with $\tau_{\mathrm{rms}} = \SI{200}{\ns}$ ($\tau_{\max} = \SI{1.93}{\us}$).}
\label{fig:snr_mcs7}
\vspace{-3mm}
\end{figure}

We then vary the received $E_b/N_0$ at a UE speed of \SI{300}{\km\per\hour} ($f_D = \SI{972}{\Hz}$).
Fig.~\ref{fig:snr} shows the impact of this on the link performance. As expected, at low received $E_b/N_0$, the link struggles with every method, including pilot transmission schemes. At such low received $E_b/N_0$, any receiver that requests pilots based on the failure of TBs will keep doing so, resulting in a high pilot overhead. This usually calls for MCS adaptation of the link, the result of which can be observed in Fig.~\ref{fig:snr_mcs7}.

From \SI{15}{\dB}, however, both DMRS receivers are limited in performance by their pilot overhead. While all learned approaches reduce their pilot transmissions as the received $E_b/N_0$ increases, CNN-ProbSparse still needs $25$ to $26$ additional bursts per test instance at \SI{20}{\dB}, where it decodes \SI{15.037}{\mega\bit}. The P2P Transformer needs $31$ to $32$ additional bursts per test instance there, and decodes \SI{14.727}{\mega\bit}. In contrast, the combination of our Multi-Subcarrier Predictor and Scaler needs only $0.33$ additional bursts per test instance on average at \SI{15}{\dB} and none at \SI{20}{\dB}, and decodes $16.339$ and \SI{16.356}{\mega\bit}, respectively. The share of lost slots that the Scaler rescues grows with the received $E_b/N_0$, from $1$ of $142$ at \SI{0}{\dB} to $216$ of $260$ at \SI{10}{\dB}, where it improves the throughput by $9.4$\%.

\subsection{Impact of the Delay Spread}
\label{subsec:delay}

\begin{figure}[t]
\centering
\includegraphics[width=\linewidth]{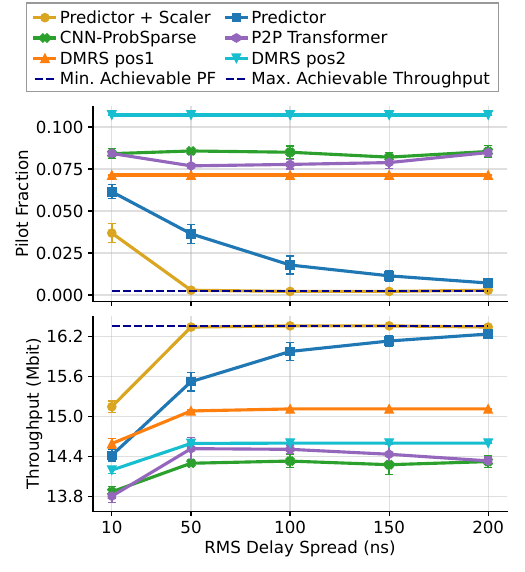}
\vspace{-5mm}
\caption{Pilot fraction (top) and correctly decoded data bits in \SI{0.5}{\s} (bottom) against the RMS delay spread $\tau_{\mathrm{rms}}$ at UE speed \SI{300}{\km\per\hour} and a received $E_b/N_0$ of \SI{15}{\dB} using MCS~$13$, for the TDL-A channel model.}
\label{fig:delay_mcs13}
\vspace{-3mm}
\end{figure}

\begin{figure}[t]
\centering
\includegraphics[width=\linewidth]{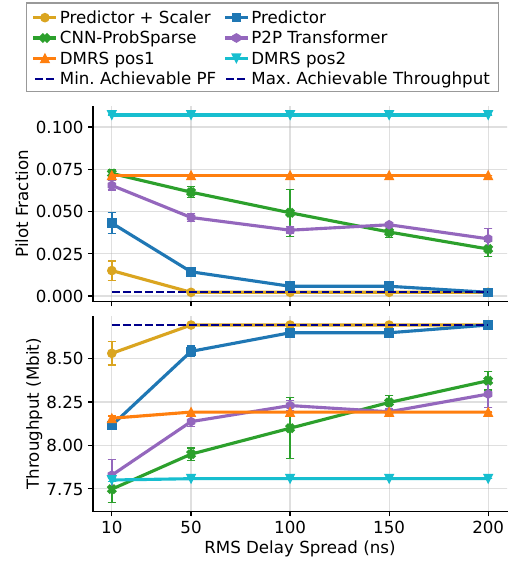}
\vspace{-5mm}
\caption{Pilot fraction (top) and correctly decoded data bits in \SI{0.5}{\s} (bottom) against the RMS delay spread $\tau_{\mathrm{rms}}$ at UE speed \SI{300}{\km\per\hour} and a received $E_b/N_0$ of \SI{15}{\dB} using MCS~$7$, for the TDL-A channel model.}
\label{fig:delay_mcs7}
\vspace{-3mm}
\end{figure}

We now vary the RMS delay spread, and with it the frequency selectivity of the channel. Figs.~\ref{fig:delay_mcs13} and~\ref{fig:delay_mcs7} show its impact for both MCSs at a UE speed of \SI{300}{\km\per\hour}. At \SI{10}{\ns}, the coherence bandwidth $B_c \approx 1/(5\tau_{\mathrm{rms}}) = \SI{20}{\MHz}$ spans the whole band, limiting the frequency diversity that LDPC can take advantage of. This results in nearly all subcarriers entering a deep fade at the same time, damaging channel estimation and refinement. With MCS~$13$, every receiver therefore decodes the fewest data bits at \SI{10}{\ns}. With a larger delay spread, as expected in urban environments with more scatterers, the data-driven receivers take advantage of the frequency diversity that a smaller coherence bandwidth brings. At MCS~$13$, CNN-ProbSparse and the P2P Transformer rise from $13.871$ and \SI{13.805}{\mega\bit} at \SI{10}{\ns} to $14.323$ and \SI{14.334}{\mega\bit} at \SI{200}{\ns}, respectively. In this range, the Multi-Subcarrier Predictor goes from $14.417$ to \SI{16.234}{\mega\bit}, a $12.6$\% improvement. At MCS~$7$, the Multi-Subcarrier Predictor reaches the maximum achievable throughput (\SI{8.694}{\mega\bit}) at \SI{200}{\ns}, and the Multi-Subcarrier Predictor + Scaler does so from \SI{50}{\ns} onward.

\subsection{Robustness to Unseen Channel Models}
\label{subsec:tdld}

While all learned models were trained on TDL-A only, we consider their performance in an alternative channel model, TDL-D, which adds a strong line-of-sight tap. This maintains the received signal power even when the scattered taps fade, thereby reducing the impact of deep fades.
We simulate TDL-D with $\tau_{\mathrm{rms}} = \SI{100}{\ns}$ ($\tau_{\max} = \SI{1.25}{\us}$) and consider all our receivers at a received $E_b/N_0$ of \SI{15}{\dB} using MCS~$13$, at UE speeds of $100$ and \SI{400}{\km\per\hour}. The results indicate that our Multi-Subcarrier Predictor is still competitive without any retraining, with its first estimate having a median NMSE of \SI{-26.2}{\dB} at \SI{100}{\km\per\hour} ($f_D = \SI{324}{\Hz}$) against \SI{-22.6}{\dB} for CNN-ProbSparse and \SI{-21.5}{\dB} for the P2P Transformer. The Multi-Subcarrier Predictor + Scaler needs only the initial burst on TDL-D as well and decodes \SI{16.356}{\mega\bit} at both speeds. However, the first estimate quality of our Multi-Subcarrier Predictor alone degrades to \SI{-13.0}{\dB}, against \SI{-20.5}{\dB} for the CNN-ProbSparse model and \SI{-18.9}{\dB} for the P2P Transformer model at \SI{400}{\km\per\hour}. CNN-ProbSparse performs nearly pilot-free channel tracking at \SI{400}{\km\per\hour}, which our Multi-Subcarrier Predictor struggles with. Since our Multi-Subcarrier Predictor tracks TDL-D at \SI{100}{\km\per\hour} accurately, and its first estimate on TDL-A at \SI{400}{\km\per\hour} still has a median NMSE of \SI{-20.1}{\dB}, we attribute this to the combination of the strong line-of-sight tap with fast-fading scattered taps, a mixture that never occurs in the training data.

\subsection{Ablation Studies}
\label{subsec:ablation}

\begin{figure}[t]
\centering
\includegraphics[width=\linewidth]{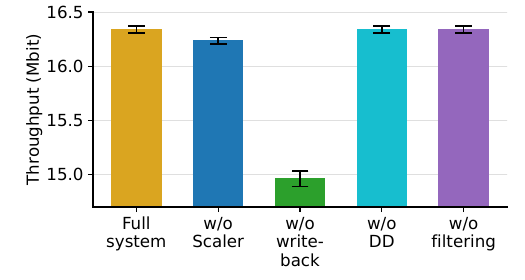}
\vspace{-2mm}
\caption{Correctly decoded data bits in \SI{0.5}{\s} of each ablation variant at UE speed \SI{300}{\km\per\hour} and a received $E_b/N_0$ of \SI{15}{\dB} using MCS~$13$, for the TDL-A channel model with $\tau_{\mathrm{rms}} = \SI{200}{\ns}$ ($\tau_{\max} = \SI{1.93}{\us}$).}
\label{fig:ablation}
\vspace{-3mm}
\end{figure}

To show what each block of Fig.~\ref{fig:pipeline} contributes, we remove one block at a time from the full receiver and keep everything else the same. We consider one operating point where the full receiver performs well in Fig.~\ref{fig:speed_mcs13}: TDL-A with $\tau_{\mathrm{rms}} = \SI{200}{\ns}$ at a UE speed of \SI{300}{\km\per\hour}, a received $E_b/N_0$ of \SI{15}{\dB} and MCS~$13$. We consider the following ablations:
\begin{enumerate}
    \item {\em Full system}: the Multi-Subcarrier Predictor + Scaler of Section~\ref{sec:pipeline}.
    \item {\em Without Decision-Directed Refinement}: the filtered prediction $\hat{\mathbf{h}}^{\mathrm{pred}}_t$ is appended to $\mathcal{H}$ in place of the DD estimate, so the slot is also decoded with the filtered predictions. The DD estimation inside the Scaler is kept.
    \item {\em Without Decoder-Aided Refinement} (``w/o write-back'' in Fig.~\ref{fig:ablation}): a decoded slot is not used to refine $\mathcal{H}$ with its re-encoded QAM symbols. The history keeps the DD estimates appended during the tracking.
    \item {\em Without filtering}: the slot is decoded with the raw DD estimates instead of their filtered version $\mathcal{L}_{\mathcal{C}}\{\hat{H}_{\mathrm{DD}}\}$.
    \item {\em Without the Scaler}: the Multi-Subcarrier Predictor alone, \ie, without the prediction scaling of Section~\ref{subsec:rescue}.
\end{enumerate}
Fig.~\ref{fig:ablation} shows the results. Without LDPC decoder-aided refinement of the channel history, the receiver loses $27$ slots per test instance on average and needs a PF of $0.0592$, against $0.33$ lost slots and $0.0029$ for the full system, and it decodes \SI{14.959}{\mega\bit} instead of \SI{16.339}{\mega\bit}. This indicates that LDPC plays a significant role in mitigating error propagation in the channel history $\mathcal{H}$ of our tracker. Similarly, the absence of DD refinement drastically lowers the first-attempt decoding rate, to only $8.5$\% of all slots. However, the soft re-estimation of~\eqref{eq:soft} and the Scaler, which both still estimate the channel from the data symbols, decode nearly all of the remaining slots, so that the receiver decodes the same TBs as the full system. In the filtering ablation, the slots that fail the first decoding attempt are decoded correctly by the soft re-estimation and the Scaler. Without the Scaler, the receiver loses $2.33$ slots on average and needs a PF of $0.0071$, which highlights the role of the Scaler in correcting Multi-Subcarrier Predictor errors.

\section{Conclusion}
\label{sec:conclusion}
In this paper, we proposed a channel tracking receiver for 5G NR PUSCH, to avoid transmitting pilots beyond an initial pilot burst. We addressed the pilot overhead in high-mobility wireless channels via channel prediction and iterative refinement. On a single subcarrier with BLER-based decisions, our proposed approach increases the throughput by $14.56$\% in comparison to pilot transmission with $16$-QAM at UE speeds up to \SI{386}{\km\per\hour} ($f_D = \SI{1250}{\Hz}$). In a more realistic 5G NR-based setting, results demonstrate that our data-driven system tracks fast-fading channels accurately with minimal pilot transmissions. At a received $E_b/N_0$ of \SI{15}{\dB}, the proposed approach nearly always requires a single burst at UE speeds up to \SI{400}{\km\per\hour} ($f_D = \SI{1296}{\Hz}$). At this speed, its PF is $97.0$\% lower than that of DMRS pos1. Its throughput is $9.1$\% higher than that of DMRS pos1 with MCS~$13$. With MCS~$7$, the throughput gain is $6.2$\%.

\bibliographystyle{IEEEtran}
\bibliography{Journal_references}

\begin{thebibliography}{10}
\providecommand{\url}[1]{#1}
\csname url@samestyle\endcsname
\providecommand{\newblock}{\relax}
\providecommand{\bibinfo}[2]{#2}
\providecommand{\BIBentrySTDinterwordspacing}{\spaceskip=0pt\relax}
\providecommand{\BIBentryALTinterwordstretchfactor}{4}
\providecommand{\BIBentryALTinterwordspacing}{\spaceskip=\fontdimen2\font plus
\BIBentryALTinterwordstretchfactor\fontdimen3\font minus \fontdimen4\font\relax}
\providecommand{\BIBforeignlanguage}[2]{{%
\expandafter\ifx\csname l@#1\endcsname\relax
\typeout{** WARNING: IEEEtran.bst: No hyphenation pattern has been}%
\typeout{** loaded for the language `#1'. Using the pattern for}%
\typeout{** the default language instead.}%
\else
\language=\csname l@#1\endcsname
\fi
#2}}
\providecommand{\BIBdecl}{\relax}
\BIBdecl

\bibitem{shannon_mathematical_nodate}
C.~E. Shannon, ``A mathematical theory of communication,'' \emph{Bell System Technical Journal}, vol.~27, no.~3, pp. 379--423, Jul. 1948.

\bibitem{ts38211}
{3GPP}, ``{NR}; physical channels and modulation,'' 3rd Generation Partnership Project (3GPP), Sophia Antipolis, France, Tech. Spec. (TS) 38.211, Release 18, V18.10.0, Sep. 2026.

\bibitem{oshea_introduction_2017}
T.~O'Shea and J.~Hoydis, ``An {Introduction} to {Deep} {Learning} for the {Physical} {Layer},'' \emph{IEEE Transactions on Cognitive Communications and Networking}, vol.~3, no.~4, pp. 563--575, Dec. 2017.

\bibitem{dorner_deep_2018}
S.~Dörner, S.~Cammerer, J.~Hoydis, and S.~ten Brink, ``Deep {Learning} {Based} {Communication} {Over} the {Air},'' \emph{IEEE Journal of Selected Topics in Signal Processing}, vol.~12, no.~1, pp. 132--143, Feb. 2018.

\bibitem{hebbar_deeppolar_2024}
S.~A. Hebbar, S.~K. Ankireddy, H.~Kim, S.~Oh, and P.~Viswanath, ``{DeepPolar}: {Inventing} {Nonlinear} {Large}-{Kernel} {Polar} {Codes} via {Deep} {Learning},'' in \emph{Proceedings of the 41st International Conference on Machine Learning ({ICML})}, ser. Proceedings of Machine Learning Research, vol. 235.\hskip 1em plus 0.5em minus 0.4em\relax Vienna, Austria: PMLR, Jul. 2024, pp. 18\,133--18\,154.

\bibitem{jiang_learn_2020}
Y.~Jiang, H.~Kim, H.~Asnani, S.~Kannan, S.~Oh, and P.~Viswanath, ``{LEARN} {Codes}: {Inventing} {Low}-latency {Codes} via {Recurrent} {Neural} {Networks},'' \emph{IEEE Journal on Selected Areas in Information Theory}, vol.~1, no.~1, pp. 207--216, May 2020.

\bibitem{lu_channel_2021}
Q.~Lu, T.~Lin, and Y.~Zhu, ``Channel {Estimation} and {Hybrid} {Precoding} for {Millimeter} {Wave} {Communications}: {A} {Deep} {Learning}-{Based} {Approach},'' \emph{IEEE Access}, vol.~9, pp. 120\,924--120\,939, 2021.

\bibitem{mattu_autoencoder_2020}
S.~R. Mattu, T.~Lakshmi~Narasimhan, and A.~Chockalingam, ``Autoencoder based {Robust} {Transceivers} for {Fading} {Channels} using {Deep} {Neural} {Networks},'' in \emph{2020 {IEEE} 91st {Vehicular} {Technology} {Conference} ({VTC2020}-{Spring})}.\hskip 1em plus 0.5em minus 0.4em\relax Antwerp, Belgium: IEEE, May 2020, pp. 1--5.

\bibitem{tr38843}
{3GPP}, ``Study on artificial intelligence ({AI})/machine learning ({ML}) for {NR} air interface,'' 3rd Generation Partnership Project (3GPP), Sophia Antipolis, France, Tech. Rep. (TR) 38.843, Release 18, V18.0.0, Dec. 2023.

\bibitem{ts38212}
{3GPP}, ``{NR}; multiplexing and channel coding,'' 3rd Generation Partnership Project (3GPP), Sophia Antipolis, France, Tech. Spec. (TS) 38.212, Release 18, V18.8.0, Dec. 2025.

\bibitem{ts38214}
{3GPP}, ``{NR}; physical layer procedures for data,'' 3rd Generation Partnership Project (3GPP), Sophia Antipolis, France, Tech. Spec. (TS) 38.214, Release 18, V18.11.0, Sep. 2026.

\bibitem{chen_channel_2021}
Z.~Chen, X.~Gao, W.~Liu, D.~Yu, and G.~Yue, ``Channel prediction for millimeter-wave {V2V} communication using autoregressive models,'' in \emph{2021 13th International Symposium on Antennas, Propagation and {EM} Theory ({ISAPE})}, Zhuhai, China, Dec. 2021, pp. 1--3.

\bibitem{jarinova_autoregressive_2013}
D.~Jarinová, ``On autoregressive model order for long-range prediction of fast fading wireless channel,'' \emph{Telecommunication Systems}, vol.~52, no.~3, pp. 1533--1539, Mar. 2013.

\bibitem{duel-hallen_fading_2007}
A.~Duel-Hallen, ``Fading {Channel} {Prediction} for {Mobile} {Radio} {Adaptive} {Transmission} {Systems},'' \emph{Proc. IEEE}, vol.~95, no.~12, pp. 2299--2313, Dec. 2007.

\bibitem{wiener_massive_mimo}
D.~L{\"o}schenbrand, M.~Hofer, and T.~Zemen, ``Spectral efficiency of time-variant massive {MIMO} using {Wiener} prediction,'' \emph{IEEE Communications Letters}, vol.~27, no.~4, pp. 1225--1229, Apr. 2023.

\bibitem{andersen_prediction_1999}
J.~B. Andersen, J.~Jensen, S.~H. Jensen, and F.~Frederiksen, ``Prediction of future fading based on past measurements,'' in \emph{Gateway to 21st {Century} {Communications} {Village}. {VTC} 1999-{Fall}. {IEEE} {VTS} 50th {Vehicular} {Technology} {Conference} ({Cat}. {No}.{99CH36324})}, vol.~1, Sep. 1999, pp. 151--155.

\bibitem{zhang_airair_2018}
Y.~Zhang, J.~Wen, G.~Yang, Z.~He, and X.~Luo, ``Air-to-air path loss prediction based on machine learning methods in urban environments,'' \emph{Wireless Communications and Mobile Computing}, vol. 2018, no.~1, 2018, art. no. 8489326.

\bibitem{gao_fast_2023}
J.~Gao, F.~Ke, C.~Zeng, and Y.~L. Lee, ``Fast time-varying wireless channel prediction based on deep learning,'' in \emph{2023 9th International Conference on Computer and Communications ({ICCC})}, Chengdu, China, Dec. 2023, pp. 940--945.

\bibitem{stenhammar_comparison_2024}
O.~Stenhammar, G.~Fodor, and C.~Fischione, ``A comparison of neural networks for wireless channel prediction,'' \emph{IEEE Wireless Communications}, vol.~31, no.~3, pp. 235--241, Jun. 2024.

\bibitem{mattu_deep_2022}
S.~R. Mattu, L.~N. Theagarajan, and A.~Chockalingam, ``Deep channel prediction: a {DNN} framework for receiver design in time-varying fading channels,'' \emph{IEEE Transactions on Vehicular Technology}, vol.~71, no.~6, pp. 6439--6453, Jun. 2022.

\bibitem{zhou_deep-learning-based_2022}
T.~Zhou, H.~Zhang, B.~Ai, C.~Xue, and L.~Liu, ``Deep-learning-based spatial–temporal channel prediction for smart high-speed railway communication networks,'' \emph{IEEE Transactions on Wireless Communications}, vol.~21, no.~7, pp. 5333--5345, Jul. 2022.

\bibitem{vaswani_attention_2017}
A.~Vaswani, N.~Shazeer, N.~Parmar, J.~Uszkoreit, L.~Jones, A.~N. Gomez, {\L}.~Kaiser, and I.~Polosukhin, ``Attention is all you need,'' in \emph{Advances in Neural Information Processing Systems ({NIPS})}, vol.~30, Long Beach, CA, USA, Dec. 2017, pp. 5998--6008.

\bibitem{jiang_accurate_2022}
H.~Jiang, M.~Cui, D.~W.~K. Ng, and L.~Dai, ``Accurate channel prediction based on transformer: {Making} mobility negligible,'' \emph{IEEE Journal on Selected Areas in Communications}, vol.~40, no.~9, pp. 2717--2732, Sep. 2022.

\bibitem{zhou_transformer_2024}
T.~Zhou, X.~Liu, Z.~Xiang, H.~Zhang, B.~Ai, L.~Liu, and X.~Jing, ``Transformer network based channel prediction for {CSI} feedback enhancement in {AI}-native air interface,'' \emph{IEEE Transactions on Wireless Communications}, vol.~23, no.~9, pp. 11\,154--11\,167, Sep. 2024.

\bibitem{guo_channel_2025}
J.~Guo, H.~Cai, and Q.~Chen, ``Channel {State} {Prediction} {Scheme} {Based} on {Deep} {Learning},'' in \emph{2025 {IEEE} 5th {International} {Conference} on {Electronic} {Technology}, {Communication} and {Information} ({ICETCI})}, May 2025, pp. 548--555.

\bibitem{rizzello_reverse_2024}
V.~Rizzello, B.~Böck, M.~Joham, and W.~Utschick, ``Reverse {Ordering} {Techniques} for {Attention}-{Based} {Channel} {Prediction},'' \emph{IEEE Open Journal of Signal Processing}, vol.~5, pp. 248--256, 2024.

\bibitem{liu_channel_2025}
Y.~Liu and Z.~Luo, ``Channel {Prediction} {Based} on {Transformer} {Architecture} in {High} {Dynamic} {MIMO}: {Mitigating} {Noise} {Effect} and {Eliminating} {Error} {Accumulation},'' in \emph{2025 {IEEE} {Wireless} {Communications} and {Networking} {Conference} ({WCNC})}, Mar. 2025, pp. 1--6.

\bibitem{nguyen_channel_2023}
C.~Nguyen, T.~M. Hoang, and A.~A. Cheema, ``Channel {Estimation} {Using} {CNN}-{LSTM} in {RIS}-{NOMA} {Assisted} {6G} {Network},'' \emph{IEEE Transactions on Machine Learning in Communications and Networking}, vol.~1, pp. 43--60, 2023.

\bibitem{lin_generative_2025}
B.~Lin, H.~Zhang, Y.~Jiang, Y.~Wang, T.~Zhang, S.~Yan, H.~Li, Y.~Liu, and F.~Gao, ``A {Generative} {Pre}-{Trained} {Language} {Model} for {Channel} {Prediction} in {Wireless} {Communications} {Systems},'' in \emph{Proceedings of the 2025 {Conference} on {Empirical} {Methods} in {Natural} {Language} {Processing}}.\hskip 1em plus 0.5em minus 0.4em\relax Suzhou, China: Association for Computational Linguistics, Nov. 2025, pp. 13\,406--13\,419.

\bibitem{he_sensing-assisted_2025}
J.~He, Z.~Ren, J.~Yao, H.~Hu, T.~X. Han, and J.~Xu, ``Sensing-{Assisted} {Channel} {Prediction} in {Complex} {Wireless} {Environments}: {An} {LLM}-{Based} {Approach},'' \emph{IEEE Wireless Communications Letters}, vol.~14, no.~12, pp. 3857--3861, Dec. 2025.

\bibitem{stformer}
C.~Xu, S.~Li, G.~Liu, L.~Yang, T.~Bi, and T.~Jiang, ``Spatial-temporal transformer for {FDD} downlink channel prediction in high-mobility scenarios,'' \emph{IEEE Transactions on Vehicular Technology}, vol.~75, no.~7, pp. 13\,474--13\,489, Jul. 2026.

\bibitem{zhou_timefrequency_2026}
T.~Zhou, Y.~Qi, Z.~Xiang, L.~Liu, B.~Ai, and X.~Jing, ``A time-frequency joint wideband {CSI} prediction model and its validation using {5G} commercial networks,'' \emph{IEEE Transactions on Vehicular Technology}, vol.~75, no.~2, pp. 2850--2863, Feb. 2026.

\bibitem{kronformer}
L.~Lagona, C.~Bockelmann, and A.~Dekorsy, ``Transformer-based pilot-to-prediction for {MIMO-OFDM} systems,'' \emph{IEEE Wireless Communications Letters}, vol.~15, pp. 2333--2337, 2026.

\bibitem{lagona_p2p_2026}
L.~Lagona, M.~Vakilifard, C.~Bockelmann, and A.~Dekorsy, ``Transformer-based pilot-to-prediction for frequency-selective channels in {OFDM} systems,'' \emph{IEEE Wireless Communications Letters}, vol.~15, pp. 310--314, 2026.

\bibitem{e2enet}
S.~Fan, H.~Li, X.~Liang, Z.~Liu, X.~Gu, and L.~Zhang, ``{E2ENet}: An end-to-end channel prediction neural network based on uplink pilot for {FDD} systems,'' \emph{IEEE Wireless Communications Letters}, vol.~13, no.~5, pp. 1285--1289, May 2024.

\bibitem{cnnlstm_ntn}
B.~De~Filippo, C.~Amatetti, and A.~Vanelli-Coralli, ``Uplink {OFDM} channel prediction with hybrid {CNN-LSTM} for {6G} non-terrestrial networks,'' in \emph{2025 Joint European Conference on Networks and Communications \& 6G Summit (EuCNC/6G Summit)}, Poznan, Poland, Jun. 2025, pp. 7--12.

\bibitem{porto_study_2025}
R.~C.~G. Porto and R.~C. de~Lamare, ``Iterative detection, decoding and channel estimation for multiple-{RIS} assisted {MIMO} systems,'' in \emph{2025 IEEE 26th International Workshop on Signal Processing and Artificial Intelligence for Wireless Communications ({SPAWC})}, Jul. 2025, pp. 1--5.

\bibitem{mehrabi_decision_2019}
M.~Mehrabi, M.~Mohammadkarimi, M.~Ardakani, and Y.~Jing, ``Decision directed channel estimation based on deep neural network $k$-step predictor for {MIMO} communications in {5G},'' \emph{IEEE Journal on Selected Areas in Communications}, vol.~37, no.~11, pp. 2443--2456, Nov. 2019.

\bibitem{schafhuber_matz_2005}
D.~Schafhuber and G.~Matz, ``{MMSE} and adaptive prediction of time-varying channels for {OFDM} systems,'' \emph{IEEE Transactions on Wireless Communications}, vol.~4, no.~2, pp. 593--602, Mar. 2005.

\bibitem{mackay_near_1996}
D.~J.~C. MacKay and R.~M. Neal, ``Near {Shannon} limit performance of low density parity check codes,'' \emph{Electronics Letters}, vol.~32, no.~18, pp. 1645--1646, Aug. 1996.

\bibitem{tr38901}
{3GPP}, ``Study on channel model for frequencies from 0.5 to 100 {GHz},'' 3rd Generation Partnership Project (3GPP), Sophia Antipolis, France, Tech. Rep. (TR) 38.901, Release 18, V18.1.0, Dec. 2025.

\bibitem{rappaport_wireless_2002}
T.~S. Rappaport, \emph{Wireless Communications: Principles and Practice}, 2nd~ed.\hskip 1em plus 0.5em minus 0.4em\relax Upper Saddle River, NJ, USA: Prentice Hall, 2002.

\bibitem{natarajkumar_fire_2026}
V.~Natarajkumar and L.~N. Theagarajan, ``Fire and forget pilots: A transformer-based approach for dynamic wireless channel tracking,'' in \emph{Proc. IEEE Global Commun. Conf. (GLOBECOM)}, Macau, China, Dec. 2026.

\bibitem{nie_time_2023}
Y.~Nie, N.~H. Nguyen, P.~Sinthong, and J.~Kalagnanam, ``A time series is worth 64 words: {Long}-term forecasting with transformers,'' in \emph{Proc. Int. Conf. Learn. Represent. (ICLR)}, Kigali, Rwanda, May 2023.

\bibitem{hoydis_sionna_2023}
J.~Hoydis, S.~Cammerer, F.~{Ait Aoudia}, A.~Vem, N.~Binder, G.~Marcus, and A.~Keller, ``Sionna: {An} open-source library for next-generation physical layer research,'' \emph{arXiv preprint arXiv:2203.11854}, Mar. 2023.

\bibitem{bengio_curriculum_2009}
Y.~Bengio, J.~Louradour, R.~Collobert, and J.~Weston, ``Curriculum learning,'' in \emph{Proceedings of the 26th Annual International Conference on Machine Learning ({ICML})}.\hskip 1em plus 0.5em minus 0.4em\relax Montreal, QC, Canada: ACM, Jun. 2009, pp. 41--48.

\bibitem{loshchilov_decoupled_2019}
I.~Loshchilov and F.~Hutter, ``Decoupled weight decay regularization,'' in \emph{Proc. Int. Conf. Learn. Represent. (ICLR)}, New Orleans, LA, USA, May 2019.

\bibitem{dmrs}
J.~Zhuang, H.~Hou, M.~Tang, W.~Wang, S.~Jin, and V.~K.~N. Lau, ``{DMRS}-based uplink channel estimation for {MU}-{MIMO} systems with location-specific {SCSI} acquisition,'' \emph{IEEE Transactions on Communications}, vol.~74, pp. 3547--3563, 2026.

\bibitem{var}
J.~H. Stock and M.~W. Watson, ``Vector autoregressions,'' \emph{Journal of Economic Perspectives}, vol.~15, no.~4, pp. 101--115, 2001.

\bibitem{ronneberger_unet_2015}
O.~Ronneberger, P.~Fischer, and T.~Brox, ``{U-Net}: Convolutional networks for biomedical image segmentation,'' in \emph{Medical Image Computing and Computer-Assisted Intervention (MICCAI)}, ser. Lecture Notes in Computer Science, vol. 9351, 2015, pp. 234--241.

\bibitem{ross_reduction_2011}
S.~Ross, G.~J. Gordon, and J.~A. Bagnell, ``A reduction of imitation learning and structured prediction to no-regret online learning,'' in \emph{Proc. Int. Conf. Artif. Intell. Statist. (AISTATS)}, Fort Lauderdale, FL, USA, Apr. 2011, pp. 627--635.

\bibitem{kingma_adam_2015}
D.~P. Kingma and J.~Ba, ``Adam: {A} method for stochastic optimization,'' in \emph{Proc. Int. Conf. Learn. Represent. (ICLR)}, San Diego, CA, USA, May 2015.

\end{thebibliography}

\end{document}